\documentclass[12pt]{article}

\usepackage[margin=1in]{geometry}
\usepackage{setspace}
\usepackage{amsmath,amssymb,amsfonts,amsthm,mathtools}
\usepackage{booktabs,array,longtable,tabularx,threeparttable,adjustbox}
\usepackage{caption,subcaption,float,pdflscape}
\usepackage[section]{placeins}
\usepackage{graphicx}
\usepackage{tikz}
\usetikzlibrary{arrows.meta,positioning,calc}
\usepackage{microtype}
\usepackage[round,authoryear]{natbib}
\usepackage[colorlinks=true,linkcolor=blue,citecolor=blue,urlcolor=blue]{hyperref}

\renewcommand{\arraystretch}{1.12}
\newcommand{\E}{\mathbb{E}}

\newcommand{\dd}{\mathrm{d}}
\newcommand{\1}{\mathbf{1}}
\newcommand{\T}{\top}

\newcommand{\diag}{\operatorname{diag}}
\newcommand{\AbsInv}{\mathrm{AbsInv}}

\newcommand{\Scarcity}{\mathrm{Scarcity}}
\newcommand{\ThinBase}{\mathrm{ThinBase}}
\newcommand{\ForcedSale}{\mathrm{ForcedSale}}
\newcommand{\FIT}{\mathrm{FIT}}

\title{Residual Supply and the Price of Risk Absorption}
\author{Ziyao Wang\thanks{I am grateful to Professor Svetlozar T. Rachev and Professor Hongwei Mei of the Department of Mathematics and Statistics at Texas Tech University for their guidance, insightful comments, and valuable suggestions to this paper.}\\Department of Mathematics and Statistics, Texas Tech University\\\texttt{ziywang@ttu.edu}}
\date{\today}

\begin{document}
\maketitle

\begin{abstract}
When redeeming open-end funds sell and natural buyers do not step in at once, some limited-capital investor must take the other side and carry the inventory until prices recover. This paper asks what return that investor requires. A continuous-time market-clearing model delivers an expected-return restriction in which the price of residual supply depends on inventory risk, trading costs, funding frictions, and the scarcity of balance sheet available to absorb it. Mapping U.S. mutual fund flows through predetermined holdings over 2003--2024, we measure one observable component of this residual supply. Forced-sale pressure predicts actual fund selling, contemporaneous price declines, and positive returns over the following one to six months. The premium roughly doubles when market-wide absorption capacity is tight, and it concentrates in stocks with thin investor bases and limited trading capacity---precisely the cross section in which clearing the imbalance should be most costly, and a pattern that mechanical return reversal does not generate.
\end{abstract}

\noindent \textit{Keywords:} asset pricing, risk absorption, demand imbalances, mutual fund flows, institutional ownership, slow-moving capital, market clearing.

\clearpage

\section{Introduction}

Traditional asset pricing explains expected returns through covariance with marginal value. This paper asks the complementary market-clearing question. When final investors change their demands because of flows, mandates, redemptions, benchmark weights, or portfolio constraints, who holds the residual supply, and what return is required to make that position acceptable?

The mechanism is clearest during mutual fund redemptions. When investors withdraw capital, a fund sells shares, and if long-horizon buyers do not absorb the order at once, the position passes to some other investor who must carry the inventory---a market maker, hedge fund, active institution, arbitrageur, or other balance-sheet-constrained buyer. That trade clears only at a price that compensates the buyer for bearing inventory risk, for financing and capital usage, and for costly execution. The shares that redeeming funds sell must be held by someone, and that someone may demand a price concession.

The model turns this intuition into a pricing restriction. Let \(\bar S_t\) and \(\bar D(P_t,X_t,Z_t)\) denote raw asset supply and final demand in share units, and let \(K_t^A\) denote the dollar capital used to scale absorbing capacity. The key market-clearing object is the required normalized residual exposure
\begin{equation}
\theta_t^{req}=\frac{1}{K_t^A}\diag(P_t)\left[\bar S_t-\bar D(P_t,X_t,Z_t)\right].
\end{equation}
This object measures residual dollars that market clearing places with risk absorbers per dollar of absorbing capital. In the absorber's local control problem, the corresponding controlled inventory is \(\theta_t^A\); equilibrium sets \(\theta_t^A=\theta_t^{req}\). The normalization is important because expected returns are rates, while raw shares are not. Section \ref{sec:model} shows that the absorber's Euler equation has the following economic content:
\[
\text{expected return}
=
\text{inventory risk compensation}
+
\text{capital and adjustment wedges}.
\]
The model does not replace stochastic discount factor pricing. It explains why risk prices and wedges can move with market-clearing pressure.

We test the model using one observable component of residual supply: open-end mutual fund flows mapped through predetermined fund holdings. The setting is useful because investor flows move capital into and out of funds, while lagged holdings reveal which stocks are mechanically exposed to those flows. Using U.S. common stocks from January 2003 through November 2024, we treat the negative part of stock-level flow-induced pressure as forced selling. Cumulative forced selling proxies for residual inventory that has not yet been fully absorbed. Form 13F ownership data measure the thickness of the natural investor base, while CRSP liquidity variables and aggregate mutual fund stress measure absorption capacity.

The evidence follows the market-clearing chain and then turns on its central link. Flow-holdings pressure predicts actual mutual fund selling; that selling depresses contemporaneous prices; and subsequent returns are positive over one- to six-month horizons. In the no-microcap sample, moving from the bottom to the top of the forced-sale rank predicts about 65 basis points over the next month and 217 basis points over six months, after standard controls. The decisive evidence, however, is where this premium concentrates. The return relation is strongest when market-wide absorption capacity is tight, when investor bases are thin, and among fund-owned stocks with limited trading capacity. This state dependence is what separates risk absorption from a generic reversal: a reversal can explain some rebound after selling, but it does not predict that the same selling commands a larger premium precisely when the market has less capacity to absorb it.

Identification rests on the timing and the cross section of the effect rather than on a single randomized shock. Forced-sale stocks do have poor pre-event returns, so the analysis isolates the part of the return relation that a mechanical loser rebound cannot produce. Actual-selling first stages, reporting-lag diagnostics, pretrend and residualized-pressure controls, matched event-time analysis, fixed effects, out-of-sample flow forecasts, ownership migration, and placebo rematching all point to the same demand-pressure channel. The object of study is how one observable component of residual supply is priced, and how that price varies across states; the portfolio evidence serves to locate the premium rather than to propose a large-capacity arbitrage strategy.

The paper connects four literatures around a single incremental claim: the price of flow-induced residual supply varies systematically with the cost of absorbing it. Demand-system and inelastic-markets models---\citet{KoijenYogo2019}, \citet{KoijenRichmondYogo2024}, and \citet{GabaixKoijen2021}---establish that investor demand moves prices, and we carry that demand imbalance into an expected-return restriction. Intermediary and slow-moving-capital theories---\citet{HeKrishnamurthy2013}, \citet{AdrianEtulaMuir2014}, \citet{ShleiferVishny1997}, and \citet{Duffie2010}---establish that scarce capital affects risk premia, and we use holdings and flows to identify which stocks are pushed toward that scarce capital. The mutual fund flow literature---\citet{CovalStafford2007}, \citet{FrazziniLamont2008}, and \citet{Lou2012}---supplies the measurement logic and documents that flow-induced selling depresses prices and reverses. Relative to this work, our contribution is to show that the reversal is not uniform: the same selling commands a larger premium exactly when balance sheet, natural demand, and trading capacity are scarce, the comparative static that liquidity and trading-cost models---\citet{PastorStambaugh2003}, \citet{AcharyaPedersen2005}, and \citet{GarleanuPedersen2013}---imply once liquidity is priced through the cost of clearing a particular residual position.

The rest of the paper proceeds as follows. Section \ref{sec:model} develops the continuous-time framework. Section \ref{sec:data} describes the data and empirical measurement. Section \ref{sec:results} presents the main results. Section \ref{sec:mechanisms} studies mechanisms and identification. Section \ref{sec:interpretation} discusses interpretation and scope. Section \ref{sec:conclusion} concludes.

\section{A Market-Clearing Model of Risk Absorption}
\label{sec:model}

\subsection{Assets, no-arbitrage, and the market-clearing object}

The model is developed locally, around the observed market state, which is the natural reference point for the comparative statics the empirical work uses. The aim is the pricing restriction that holds when residual supply is absorbed by investors with convex costs; the tests then exploit the signs, interactions, and timing implied by the resulting Euler equation rather than estimates of the primitive cost parameters.

The argument rests on four local conditions. First, final demand is differentiable near the observed state and not infinitely price elastic. Second, price-response signs are imposed in normalized dollar-exposure units rather than inferred from raw share demand, so raw quantities are converted into dollars of exposure per dollar of absorbing capital before entering the absorber's problem. Third, the absorption cost is convex and smooth within a region of fixed position signs and a fixed capital-constraint regime, with marginal costs at kinks read as subgradients or as limits of smooth approximations. Fourth, the absorber adjusts the finite-variation component of exposure on an interior margin, while discontinuous demand shocks are treated as block imbalances. Together these conditions are enough to deliver the pricing equation, its SDF interpretation, and the discrete-time stock-flow approximation used in the tests.

A one-period benchmark gives the comparative statics before the continuous-time notation is introduced. Let final normalized dollar demand be \(n(P,Z)\), normalized supply be \(s\), and residual exposure be
\[
\theta=s-n(P,Z).
\]
Suppose the absorber's local cost is \(C(\theta;\kappa)=\frac{1}{2}a(\kappa)\theta^2\), where \(\kappa\) is absorption scarcity and \(a'(\kappa)>0\). The clearing return must satisfy
\[
\pi=\frac{\partial C(\theta;\kappa)}{\partial \theta}=a(\kappa)\theta.
\]
Thus \(\partial \pi/\partial \theta=a(\kappa)>0\) and \(\partial^2 \pi/(\partial\theta\,\partial\kappa)=a'(\kappa)>0\). More residual supply requires a higher return, and the same residual supply is more expensive when absorption capacity is scarce. The continuous-time model below adds covariance across assets, costly adjustment, funding wedges, and future-state risks, but the empirical tests are built around these two signs.

Consider an economy with \(N\) risky assets and a locally riskless asset with rate \(r_t\). Risky total returns satisfy
\begin{equation}
\dd R_t=\mu_t\dd t+\sigma_t\dd W_t,
\label{eq:return_process}
\end{equation}
where \(W_t\) is an \(m\)-dimensional Brownian motion and \(\sigma_t\) is an \(N\times m\) volatility matrix. This return process separates compensation from risk exposure: \(\mu_t\) is the conditional mean return that prices must deliver, while \(\sigma_t\dd W_t\) is the instantaneous risk that investors must hold. The vector of conditional expected excess returns is \(\pi_t=\mu_t-r_t\1\). If a stochastic discount factor \(M_t\) exists with diffusion
\begin{equation}
\frac{\dd M_t}{M_t}=-r_t\dd t-\lambda_t^{\T}\dd W_t,
\label{eq:sdf_diffusion}
\end{equation}
the vector \(\lambda_t\) prices the Brownian risks in the same probability space as returns. A high component of \(\lambda_t\) means that the marginal investor assigns a high shadow value to exposure to that shock. In a frictionless and complete market this gives \(\pi_t=\sigma_t\lambda_t\). With trading costs, segmentation, funding wedges, or binding constraints, the pricing relation can be written more generally as
\begin{equation}
\pi_t=\sigma_t\lambda_t+\omega_t,
\label{eq:pricing_wedge}
\end{equation}
where \(\omega_t\) collects wedges on constrained or costly margins. Here \(\lambda_t\) is the \(m\)-dimensional diffusion market price of risk associated with whatever SDF supports asset prices at the aggregate equilibrium level. It is compact notation for the risk prices embedded in \(M_t\), not a new observable factor introduced by the paper. Later, when the marginal risk absorber's value process is characterized, we write the absorber-implied risk price as \(\lambda_t^A\). On unconstrained margins in which the absorber is marginal and no additional trading wedge remains, \(\lambda_t=\lambda_t^A\) and \(\pi_t=\sigma_t\lambda_t^A\). On constrained margins, the same absorber marginal value generates risk prices, but expected returns also include the wedge terms collected in \(\omega_t\). This paper explains where both the risk-price component and the wedge component come from when market clearing requires limited-capital investors to absorb residual supply.

Let final investors have raw asset demand
\begin{equation}
\bar q_t^E=\bar D(P_t,X_t,Z_t),
\label{eq:raw_demand}
\end{equation}
where \(P_t\) is the price vector, \(X_t\) contains asset characteristics and cash-flow states, and \(Z_t\) contains demand states such as flows, mandates, benchmark weights, redemptions, and allocation rules. Economically, this demand system collects all investors whose positions are not perfectly elastic arbitrage capital: their desired holdings change with prices, characteristics, and institutional demand states. Let raw net supply be \(\bar S_t\). Market clearing in raw quantities requires
\begin{equation}
\bar q_t^E+\bar q_t^A=\bar S_t,
\label{eq:raw_clearing}
\end{equation}
which states that every share must be held either by final investors or by the risk-absorbing sector. When final demand does not absorb the full supply at current prices, the remaining position is assigned to risk absorbers. The raw residual position is therefore
\begin{equation}
\bar q_t^A=\bar S_t-\bar D(P_t,X_t,Z_t).
\label{eq:raw_residual}
\end{equation}
Positive components of \(\bar q_t^A\) mean that risk absorbers must be net long the corresponding raw shares; negative components mean they must absorb net short exposure. This raw position is not the object that enters the expected-return equation, because expected returns are rates rather than dollar or share quantities. Let \(K_t^A>0\) denote the dollar capital of the relevant risk-absorbing sector used to scale positions. Define normalized supply and demand by
\begin{equation}
s_t=\frac{1}{K_t^A}\diag(P_t)\bar S_t,
\qquad
n_t(P_t,X_t,Z_t)=\frac{1}{K_t^A}\diag(P_t)\bar D(P_t,X_t,Z_t).
\label{eq:normalized_supply_demand}
\end{equation}
These transformations convert raw shares into dollars of exposure per dollar of absorbing capital. A large value of \(s_{i,t}\) or \(n_{i,t}\) therefore means that the asset represents a large balance-sheet claim relative to the capital available to absorb it. The residual exposure held by risk absorbers is
\begin{equation}
\theta_t^{req}=\frac{1}{K_t^A}\diag(P_t)\bar q_t^A
=s_t-n_t(P_t,X_t,Z_t).
\label{eq:theta}
\end{equation}
The object in equation \eqref{eq:theta} is the residual exposure required by market clearing. It is the normalized dollar exposure that the absorbing sector must hold for the market to clear. It is not yet the state equation in the absorber's dynamic program. Let \(\theta_t^A\) denote the absorber's controlled inventory state in the same normalized exposure units. Market clearing imposes the equilibrium restriction
\begin{equation}
\theta_t^A=\theta_t^{req}.
\label{eq:theta_equilibrium}
\end{equation}
After this restriction is imposed, we suppress the superscripts and write the common exposure as \(\theta_t\). The unit of both \(\theta_t^{req}\) and \(\theta_t^A\) is dollars of residual exposure per dollar of relevant risk-absorbing capital. This scaling matters because expected returns are rates, while the marginal cost of holding a raw share position is not. With the dollar-exposure scaling, \((\theta_t^A)^{\T}\pi_t\) is an excess return on absorber capital and the model's marginal conditions map directly into expected-return units. The capital state \(K_t\) used below in the cost function is a normalized balance-sheet capacity state; the dollar capital \(K_t^A\) in equation \eqref{eq:theta} is the scaling variable that converts raw positions into that common exposure space.

\begin{table}[t]
\centering
\caption{Notation in the Market-Clearing and HJB Blocks}
\label{tab:theta_notation}
\small
\begin{tabularx}{0.96\textwidth}{@{}lX@{}}
\toprule
Symbol & Meaning \\
\midrule
\(\bar q_t^A\) & Raw residual shares left for risk absorbers by the raw clearing equation. \\
\(\theta_t^{req}\) & Market-clearing required residual exposure, measured as normalized dollar exposure. \\
\(\theta_t^A\) & Absorber inventory state controlled in the local HJB. \\
\(u_t\) & Trading speed, with controlled law \(\dd\theta_t^A=u_t\dd t\). \\
\(\mathcal Y_t\) & State vector \((Z_t,K_t,\Sigma_t,\Phi_t,\Psi_t)\). \\
\(\mathcal L^Y\) & Generator of \(\mathcal Y_t\), acting only on the exogenous state. \\
\(\mathcal A^u\) & Controlled generator of the full state \((\theta^A,\mathcal Y)\). \\
\(p_t\) & Inventory shadow value, \(p_t=V_\theta=\nabla_u c\) at the optimum. \\
\bottomrule
\end{tabularx}
\end{table}

The dynamic program below should be read together with this clearing condition, since risk absorbers do not choose an exposure unrelated to market clearing. Given prices and expected returns, they select a locally optimal adjustment path, and in equilibrium prices and expected returns must make that path consistent with the normalized residual supply assigned to them by \(s_t-n_t(P_t,X_t,Z_t)\). Equivalently, the final-demand system determines how much normalized supply is left for risk absorbers, while the absorber's Euler equation determines the expected return required to hold it.

A single-asset example fixes the economics. Suppose an open-end fund complex must sell shares after redemptions. If long-horizon natural buyers absorb only part of the order at current prices, the remaining dollar exposure passes to a market maker, hedge fund, active institution, or other risk absorber. The expected return required to hold it is low when the stock is liquid, uncorrelated with the absorber's existing inventory, and cheap to finance, and high when the stock adds to an already crowded inventory portfolio, trades with large price impact, or consumes scarce balance-sheet capacity. The model below formalizes this price concession and the subsequent expected return as the marginal cost of clearing the residual exposure.

The demand state \(Z_t\) is allowed to be stochastic and may include persistent flows or jumps. In the absorber's local control problem, the controlled state is \(\theta_t^A\). On the continuous finite-variation adjustment margin, and away from block shocks and jumps, the absorber changes inventory at trading speed \(u_t\):
\begin{equation}
\dd\theta_t^A=u_t\dd t,
\label{eq:fv_trading}
\end{equation}
where \(u_t\) is trading speed in normalized dollar-exposure units. Economically, \(u_t\) is the deliberate pace at which the absorbing sector works inventory up or down. Demand shocks, price-induced demand responses, and capital-scaling effects move the market-clearing requirement \(\theta_t^{req}\); they are connected to the absorber's controlled inventory through the equilibrium condition \(\theta_t^A=\theta_t^{req}\). If one instead models raw shares as the controlled state, the mechanically induced terms from \(P_t\) and \(K_t^A\) must be added to the state dynamics. Jump shocks can be handled as block absorption costs or event-time shocks; empirically, large mutual fund flow pressure episodes serve precisely this role.

The required residual exposure in equation \eqref{eq:theta} is the object that links demand systems to expected returns. A local expansion makes this link explicit. Around a state \((P_t,X_t,Z_t)\), using the normalized demand function \(n_t(\cdot)\),
\begin{equation}
\dd\theta_t^{req}
=\dd s_t-n_{P,t}\dd P_t-n_{X,t}\dd X_t-n_{Z,t}\dd Z_t+\text{higher-order terms}.
\label{eq:theta_expansion}
\end{equation}
This formula decomposes changes in the market-clearing residual exposure into supply shocks, price-induced demand responses, characteristic changes, and demand-state shocks. It should not be read as the controlled inventory law used in the HJB; that law in the absorber's dynamic program is \(\dd\theta_t^A=u_t\dd t\). The matrix \(n_{P,t}\) is the derivative of normalized dollar demand rather than of raw share demand, and the distinction matters. In one asset, \(n(P)=P\bar D(P)/K^A\), so
\[
n_P(P)=\frac{\bar D(P)+P\bar D'(P)}{K^A}.
\]
Thus raw share demand can be downward sloping, \(\bar D'(P)<0\), while normalized dollar demand is not downward sloping unless share demand is sufficiently elastic. Equivalently, \(n_P<0\) requires the absolute value of the raw share-demand elasticity to exceed one. In multiple assets, negative own-price derivatives of raw share demand also do not imply that \(n_{P,t}\) is negative semidefinite. Whenever the paper uses the sign of this object, it is therefore a local condition on normalized dollar demand, or more precisely on the relevant symmetric part of the price-response matrix in normalized exposure units. Under that local dollar-demand condition, a price concession raises normalized final demand and lowers the residual exposure left for risk absorbers. If normalized dollar demand is close to inelastic, shocks to supply, flows, or mandates pass through more strongly to \(\theta_t\). The matrices \(n_{Z,t}\) and \(n_{X,t}\) determine which assets are pushed toward risk absorbers when the demand state changes. The empirical flow-holdings measure is a discrete-time proxy for the \(n_{Z,t}\dd Z_t\) term. If one starts from raw share quantities, \(\dd s_t\) and \(\dd n_t\) also include the price-scaling and capital-scaling terms generated by \(\diag(P_t)/K_t^A\). Writing the clearing equation directly in normalized exposure units keeps these terms inside \(s_t\) and \(n_t\), which is the unit convention used throughout the pricing equation.

This expansion also clarifies why the theory is not a pure liquidity story. Liquidity enters through the cost of adjusting \(\theta_t\), but the amount that must be adjusted is determined by the final-demand system. Two stocks with identical bid-ask spreads can have different expected returns if one has a broad natural investor base and the other must be placed with scarce arbitrage capital after a flow shock. Conversely, two stocks with similar institutional ownership can have different absorption premia if one is much more costly to trade. The pricing object is the product of residual supply and the marginal cost of absorbing it.

The state vector is allowed to evolve with both continuous and discontinuous components. A convenient representation is
\begin{align}
\dd Z_t &= b_Z(\mathcal Y_t)\dd t+\sigma_Z(\mathcal Y_t)\dd W_t^Z+J_Z(\mathcal Y_t)\dd N_t,\label{eq:z_process}\\
\dd K_t &= b_K(\mathcal Y_t)\dd t+\sigma_K(\mathcal Y_t)\dd W_t^K,\label{eq:k_process}\\
\dd \Sigma_t &= b_\Sigma(\mathcal Y_t)\dd t+\sigma_\Sigma(\mathcal Y_t)\dd W_t^\Sigma,\label{eq:sigma_process}
\end{align}
where \(\mathcal Y_t=(Z_t,K_t,\Sigma_t,\Phi_t,\Psi_t)\). These laws of motion make demand conditions, absorbing capacity, and risk conditions state variables rather than fixed background parameters. The Brownian shocks capture ordinary flow and volatility innovations, while the jump component captures large redemptions, index events, or other block imbalances. The baseline Euler equation is written for the continuous margin. Jump or omitted future-state risks enter either through the extended hedging term \(h_t\) or through a separate block absorption cost in the appendix. This distinction keeps the continuous-time notation rigorous: \(u_t\) is a controlled trading speed, not the derivative of an arbitrary semimartingale demand shock.

\subsection{Risk absorber costs and the Euler equation}

Risk absorbers earn excess returns from holding residual exposure but incur costs. Their inventory risk cost is
\begin{equation}
\frac{1}{2}\gamma_t\theta_t^{\T}\Sigma_t\theta_t,
\qquad \Sigma_t=\sigma_t\sigma_t^{\T},
\label{eq:inventory_cost}
\end{equation}
where \(\gamma_t\) is effective risk aversion, or equivalently the scarcity of risk-bearing capacity. This cost is large when the residual portfolio is volatile or its positions covary strongly with one another, and it rises as risk-bearing capacity becomes scarce. Trading adjustment costs are
\begin{equation}
\frac{1}{2}u_t^{\T}\Phi_tu_t,
\label{eq:trading_cost}
\end{equation}
with \(\Phi_t\) large when trading capacity is low, capturing the idea that absorbing a flow shock quickly is more expensive than holding the same position once it has been worked into the market. Funding and balance-sheet costs are summarized by \(\ell_t(\theta_t,K_t)\), where \(K_t\) is the state of absorbing capital. A useful example is
\begin{equation}
\ell_t(\theta,K)=\frac{1}{2}\theta^{\T}\Psi_t\theta+\frac{\eta_t}{2}\left[\max\{0,m_t^{\T}|\theta|-K_t\}\right]^2.
\label{eq:capital_cost}
\end{equation}
The first term captures margin, financing, and balance-sheet haircuts, while the second captures the nonlinear increase in marginal cost as required capital approaches available capacity. The example is convex and piecewise smooth, and the Euler equations below are stated on regions with fixed position signs and a fixed slack-or-binding capital regime; at kink points, gradients are read as subgradients, or equivalently one may replace \(|\theta|\) and the positive-part operator by smooth approximations without changing the local comparative statics.

The cost function is best read as a local second-order approximation to a more general risk-management problem: the inventory term is the certainty-equivalent cost of holding a portfolio with instantaneous variance \(\theta_t^{\T}\Sigma_t\theta_t\), the trading term is the price impact or implementation shortfall of changing exposure at speed \(u_t\), and the capital term reflects that the same risk exposure is more expensive when it consumes scarce balance sheet or collateral. Convexity is the key restriction, ensuring that absorbers require higher compensation for larger positions and that the marginal price of absorption rises when positions are crowded or capital is scarce.

Before introducing dynamic adjustment, the static intuition is immediate. If trading costs and separate hedging motives are absent, the local first-order condition is
\begin{equation}
\pi_t=\gamma_t\Sigma_t\theta_t+\nabla_\theta\ell_t(\theta_t,K_t).
\label{eq:static_absorption}
\end{equation}
The stock must offer a higher expected return when it loads positively on the absorber's inventory portfolio or consumes scarce balance-sheet capacity. Dynamic trading adds a costate term because the absorber cannot instantaneously move inventory without paying adjustment costs.

Risk absorbers choose trading speed to maximize the discounted value of returns net of absorption costs,
\begin{equation}
\max_{\{u_s\}}\E_t\int_t^\infty e^{-\rho(s-t)}\left[(\theta_s^A)^{\T}\pi_s-c_s(\theta_s^A,u_s,K_s)\right]\dd s,
\label{eq:absorber_objective}
\end{equation}
which says that the absorbing sector values the expected excess return from holding residual supply but subtracts the costs of risk, trading, and balance-sheet usage. The discount rate \(\rho\) summarizes how the sector trades off current compensation against future adjustment costs. The instantaneous cost rate is
\begin{equation}
 c_t(\theta,u,K)=\frac{1}{2}\gamma_t\theta^{\T}\Sigma_t\theta+\frac{1}{2}u^{\T}\Phi_tu+\ell_t(\theta,K).
\label{eq:cost_function}
\end{equation}
This cost rate bundles the three economic frictions that make absorption costly: bearing inventory risk, trading into or out of the position, and using scarce financing or capital capacity. Because the benefit term \((\theta_s^A)^{\T}\pi_s\) and the cost term \(c_s\) are both measured per dollar of absorbing capital, the first-order condition delivers expected returns in rate units.

For the local HJB derivation, the state variable is the absorber's controlled inventory \(\theta^A\), and the remaining state vector is \(Y_t\equiv\mathcal Y_t=(Z_t,K_t,\Sigma_t,\Phi_t,\Psi_t)\). Let \(\mathcal L^Y\) be the generator of \(Y_t\). We assume that \(V(\theta,y)\) is sufficiently smooth on the local region considered, that the optimal trading speed is interior, and that the coefficients of \(\mathcal L^Y\) depend on \(Y\) but not directly on the individual absorber's controlled inventory \(\theta^A\). Under these conditions,
\begin{equation}
\nabla_\theta(\mathcal L^YV)=\mathcal L^Y(V_\theta).
\label{eq:generator_exchange}
\end{equation}
The absorber is price-taking in this local problem: \(\pi\) is held fixed when differentiating the absorber's value function. Market clearing later determines the value of \(\pi\) that supports the required residual exposure.

For a smooth function \(f(\theta,y)\), define the controlled generator of the full state \((\theta^A,Y)\) by
\begin{equation}
\mathcal A^u f(\theta,y)
=u^{\T}\nabla_\theta f(\theta,y)+\mathcal L^Y f(\theta,y).
\label{eq:controlled_generator}
\end{equation}
Let \(p(\theta,y)=V_\theta(\theta,y)\) denote the shadow value of an additional unit of normalized inventory and write \(p_t=p(\theta_t^A,Y_t)\). The HJB and the optimal trading condition are
\begin{align}
\rho V(\theta,y)
&=\max_u\left\{\theta^{\T}\pi-c(\theta,u,K)+V_\theta(\theta,y)^{\T}u+\mathcal L^YV(\theta,y)\right\},
\label{eq:hjb_main}\\
V_\theta(\theta,y)&=\nabla_u c(\theta,u,K).
\label{eq:trading_foc_main}
\end{align}
Taking the envelope derivative of the HJB with respect to the controlled inventory state, using the interior first-order condition and the exchangeability condition in \eqref{eq:generator_exchange}, gives
\[
\rho V_\theta
=\pi-\nabla_\theta c+V_{\theta\theta}u+\mathcal L^Y(V_\theta)
=\pi-\nabla_\theta c+\mathcal A^u p.
\]
Rearranging and evaluating at time \(t\) gives the baseline Euler equation
\begin{equation}
\pi_t
=\nabla_\theta c_t(\theta_t^A,u_t,K_t)
-\mathcal A^{u_t}p(\theta_t^A,Y_t)
+\rho p_t,
\qquad
p_t=V_\theta(\theta_t^A,Y_t)=\nabla_uc_t(\theta_t^A,u_t,K_t).
\label{eq:euler_general}
\end{equation}
This equation is the dynamic version of the static absorption condition. The expected return must cover the current marginal holding cost \(\nabla_\theta c_t\), adjusted for the predictable drift of the inventory shadow value. Equivalently, \(\mathcal A^{u_t}p(\theta_t^A,Y_t)\) is the predictable drift of \(p_t\) under the controlled law \(\dd\theta_t^A=u_t\dd t\). The shorter notation \(D_t[p_t]\), if used informally, refers to this drift.

If the value function fully includes all future demand, capital, and opportunity-state effects, equation \eqref{eq:euler_general} already incorporates their drift through \(\mathcal A^{u_t}p\). In the main representation we allow an additional term \(h_t\) only for incremental hedging components not explicitly summarized by the current inventory and trading-cost state. Thus \(h_t\) is zero in the baseline derivation and can be written as \(h_t=h_t^{demand}+h_t^{capital}+h_t^{opportunity}\) in the extended specification. The empirical tests focus on the directly measurable inventory, flow, and scarcity components rather than estimating this residual hedging term. With quadratic holding costs and adjustment costs, the extended pricing equation is
\begin{equation}
\pi_t
=\gamma_t\Sigma_t\theta_t^A
+\nabla_\theta\ell_t(\theta_t^A,K_t)
-\mathcal A^{u_t}p(\theta_t^A,Y_t)
+\rho p_t+h_t.
\label{eq:euler_quadratic}
\end{equation}
Under the quadratic adjustment cost in \eqref{eq:trading_cost}, \(p_t=\Phi_tu_t\), but the generator in \eqref{eq:euler_quadratic} acts on the policy-induced costate function \(p(\theta,y)\), not mechanically on a realized product \(\Phi_tu_t\). After imposing the equilibrium condition \(\theta_t^A=\theta_t^{req}\), the common exposure is again written as \(\theta_t\). Equation \eqref{eq:euler_quadratic} is the model's central pricing equation. The first term is compensation for covariance between each asset and the inventory portfolio. The second is compensation for funding and balance-sheet usage. The third and fourth are the shadow costs of dynamic adjustment. The final term reflects hedging demand against future states in which absorption becomes more expensive.

The hedging term \(h_t\) plays no role in the tests, which draw their content from measured absorption-pressure variables: forced residual supply should be followed by positive expected returns, and the price of that supply should rise when trading capacity, natural demand, or absorbing capital is scarce.

The Euler equation is derived on the continuous adjustment margin. Demand jumps and block imbalances move the required residual exposure \(\theta_t^{req}\) and can be handled as boundary changes, block absorption costs, or event-time shocks. They are not part of the continuous HJB drift \(\dd\theta_t^A=u_t\dd t\).

For asset \(i\), the inventory component can be written as
\begin{equation}
\left[\gamma_t\Sigma_t\theta_t\right]_i
=\gamma_t\,\frac{\dd\left\langle R_i,\int_0^\cdot \theta_{s-}^{\T}\dd R_s\right\rangle_t}{\dd t}.
\label{eq:inventory_covariance}
\end{equation}
This representation makes the economic content of the term explicit. With predictable exposure \(\theta_{t-}\) held fixed over the instant used to compute the quadratic covariation, a stock commands a high inventory premium not because its own volatility is high, but because it covaries with the absorber's total inventory portfolio: absorbing one more dollar of a stock that resembles positions the absorber already holds is costly, whereas a stock that hedges the existing inventory may require little compensation or even a negative premium. The model thus nests a covariance logic, but the relevant covariance is with the endogenous inventory that market clearing assigns to constrained investors.

The dynamic trading term has a similarly direct interpretation, but its sign depends on the direction of adjustment. The costate \(p_t=V_\theta(\theta_t^A,Y_t)\) is the shadow value of carrying an additional unit of normalized inventory. At the optimum it equals the marginal execution cost \(\Phi_tu_t\) under quadratic adjustment costs. The term \(\rho p_{i,t}\) compensates the absorber for tying up the shadow value of the inventory position, while \(-[\mathcal A^{u_t}p(\theta_t^A,Y_t)]_i\) reflects expected changes in that value. For a forced-sale episode in which the absorber carries positive residual exposure, positive subsequent returns mainly come from the current marginal holding and capital costs, and can be amplified by block absorption costs. The dynamic adjustment term can raise or lower the premium depending on the sign of \(u_t\) and the expected path of \(p_t\). This convention matters because gradual liquidation of a positive inventory position may have the opposite trading-speed sign from gradual accumulation.

The capital term is state dependent even when the physical risk of the asset is unchanged. While \(m_t^{\T}|\theta_t|<K_t\), the nonlinear constraint in \(\ell_t\) is slack and marginal capital costs are roughly linear, but as \(m_t^{\T}|\theta_t|\) approaches or exceeds \(K_t\) the marginal cost rises sharply. The same stock, with the same beta and the same flow shock, can therefore require a higher expected return in a stressed funding state than in a normal one---the theoretical basis for the empirical interaction between absorption pressure and the scarcity proxy.

The equation is compatible with stochastic discount factor pricing. If risk absorbers are marginal on an unconstrained trading margin, their marginal value defines an SDF \(M_t^A\) with
\begin{equation}
\frac{\dd M_t^A}{M_t^A}=-r_t\dd t-(\lambda_t^A)^{\T}\dd W_t.
\label{eq:absorber_sdf}
\end{equation}
This SDF is the marginal value of wealth for the sector that clears residual supply, and its diffusion risk price \(\lambda_t^A\) is high in states where an additional dollar of wealth is especially valuable to constrained absorbers. On margins without trading wedges it satisfies \(\pi_t=\sigma_t\lambda_t^A\), with the risk price decomposing into an inventory component \(\gamma_t\sigma_t^{\T}\theta_t\), a capital component, a demand-state component, and a funding component; on constrained margins, expected returns additionally carry trading and funding wedges. The theory therefore does not contradict SDF pricing but gives economic content to the risk prices and wedges.

The wedge representation can be made explicit by writing
\begin{equation}
\omega_t^{abs}
=\nabla_\theta\ell_t(\theta_t^A,K_t)
-\mathcal A^{u_t}p(\theta_t^A,Y_t)+\rho p_t+h_t^{wedge}.
\label{eq:wedge_decomposition}
\end{equation}
The wedge \(\omega_t^{abs}\) is the part of expected returns that arises because the absorber's balance sheet and trading technology are costly. It is positive when absorbing the residual position consumes scarce capital, requires costly trading, or exposes the absorber to future states in which liquidation is expensive.
Then
\begin{equation}
\pi_t=\sigma_t\lambda_t^{inv}+\omega_t^{abs},
\qquad
\lambda_t^{inv}=\gamma_t\sigma_t^{\T}\theta_t,
\label{eq:inventory_risk_price}
\end{equation}
which separates the covariance price of inventory risk from the nonfrictionless absorption wedge. The term \(\lambda_t^{inv}\) is the market price of Brownian risk generated by the absorber's inventory portfolio; \(\omega_t^{abs}\) collects the costs that cannot be represented as a frictionless covariance price alone. Up to additional diffusion-risk prices contained in \(h_t\), the SDF language and the market-clearing language are therefore complementary. The SDF representation states that expected returns are priced by marginal value. The market-clearing representation explains why the marginal value loads on inventory, capital, and demand states.

\subsection{Stock-flow representation and testable implications}

Equation \eqref{eq:euler_quadratic} is not taken to the data as a structural estimating equation. Near a stable equilibrium with locally linear demand and locally quadratic costs, it motivates a first-order empirical projection:
\begin{equation}
\pi_t=A_t\theta_t+B_t\E_t\left[\frac{\dd\theta_t^{res}}{\dd t}\right]+C_tu_t+G_t\kappa_t+H_t(\theta_t\otimes\kappa_t)+\varepsilon_t.
\label{eq:stock_flow}
\end{equation}
Here \(\theta_t\) is residual inventory, \(\E_t[\dd\theta_t^{res}/\dd t]\) is expected residual flow pressure, \(u_t\) is trading speed, and \(\kappa_t\) is absorption-capacity scarcity. The coefficients collect local derivatives of inventory risk, capital costs, trading shadow costs, and hedging components; they are not unique structural estimates of \(\gamma_t\), \(\Phi_t\), or \(\ell_t\). The interaction term expresses a key comparative static: the same residual inventory is more costly when risk-bearing capacity is scarce.

The stock-flow specification is the discrete-time empirical approximation motivated by the local Euler equation. In monthly data, the econometrician observes flow-induced pressure, lagged holdings, and future returns rather than the instantaneous process \(\theta_t\). If residual inventory follows
\begin{equation}
\theta_{i,t+\Delta}=(1-\delta_i\Delta)\theta_{i,t}+\Delta\theta_{i,t}^{res}+o(\Delta),
\label{eq:inventory_decay}
\end{equation}
then a decaying sum of past forced sales is a natural proxy for \(\theta_{i,t}\). If fund flows are persistent, \(\E_t[\Delta\theta_{i,t+\Delta}^{res}]\) is nonzero and should also be priced. If the trading or capital state changes over time, the price of both stock and flow pressure changes with \(\kappa_t\). This is why the empirical design includes current forced sales, \(\AbsInv\), expected future sales, and interactions with scarcity and investor-base variables.

The empirical predictions follow directly. First, stocks with greater forced residual supply should earn higher future returns. Second, expected future selling pressure should predict returns if investors require compensation before the flow is fully absorbed. Third, the inventory-return relation should be stronger when market liquidity is poor, when mutual fund redemption stress is high, and when a stock has a thin natural investor base. Fourth, price dynamics should show a negative contemporaneous price effect around forced selling and subsequent positive returns as the inventory is absorbed. Fifth, absorption portfolios should contain abnormal returns not spanned by standard factor models, especially outside the largest and most liquid stocks.

Several limiting cases help locate the model relative to standard theories. If final demand is perfectly elastic and risk absorbers never hold residual supply, \(\theta_t=0\) and the absorption premium disappears. If trading costs and capital costs are zero, equation \eqref{eq:euler_quadratic} collapses to an inventory-risk pricing equation. If \(\theta_t\) is proportional to the market portfolio, the inventory term becomes a CAPM-like covariance relation. If capital costs dominate, the model resembles intermediary asset pricing. If trading costs dominate and demand shocks decay slowly, the model produces price pressure followed by reversal. These limiting cases are useful because they show that risk absorption is not a separate alternative to asset pricing theory; it is a market-clearing mechanism that selects which risks and wedges matter in equilibrium.

The tests below focus on these directly observable restrictions: residual-supply shocks should move prices when they arrive, earn compensation while they are absorbed, and command higher prices when trading capacity and natural demand are weak.

\section{Data and Empirical Measurement}
\label{sec:data}

\subsection{Stock sample and controls}

The empirical analysis uses U.S. common stocks from CRSP, the Center for Research in Security Prices, matched to Compustat characteristics and institutional ownership data. The sample starts in January 2003 because the mutual fund holdings coverage used in the analysis is cleanest from that point onward, and it ends in November 2024. Common stocks are identified using CRSP share codes 10 and 11. Returns include delisting returns when available. The main sample contains 1,022,728 stock-month observations and 10,338 distinct PERMNOs, where PERMNO is the permanent security identifier assigned by CRSP. The no-microcap sample keeps stocks with market equity above the 20th percentile of New York Stock Exchange (NYSE) market equity in the same month. It contains 490,162 stock-month observations and 5,559 PERMNOs.

The controls follow the cross-sectional asset pricing literature. They include log market equity, book-to-market, past twelve-month momentum excluding the most recent month, short-term reversal, market beta, idiosyncratic volatility, turnover, Amihud illiquidity, institutional ownership, ownership breadth, and ownership concentration. All stock-level explanatory variables are lagged relative to future returns. Continuous variables used in cross-sectional regressions are transformed into monthly ranks, which improves comparability across variables and reduces the influence of extreme observations.

\subsection{Mutual fund flows and stock-level demand pressure}

The flow of fund \(f\) in month \(t\) is constructed from total net assets and fund returns. Let \(TNA_{f,t}\) denote total net assets and \(R_{f,t}\) denote the fund's net return. The investor flow rate is
\begin{equation}
Flow_{f,t}=\frac{TNA_{f,t}-TNA_{f,t-1}(1+R_{f,t})}{TNA_{f,t-1}}.
\end{equation}
This measure removes the mechanical effect of fund returns on assets and isolates investor contributions and redemptions. The core stock-level demand-pressure measure maps fund flows through lagged holdings:
\begin{equation}
\FIT_{i,t}=\sum_f w_{f,i,t-1}Flow_{f,t}TNA_{f,t-1}.
\end{equation}
We refer to \(\FIT_{i,t}\) as flow-induced trading. It is a dollar measure in which \(Flow_{f,t}TNA_{f,t-1}\) is the dollar flow into or out of fund \(f\) and \(w_{f,i,t-1}\) allocates that flow to stock \(i\) using predetermined holdings. The construction has a shift-share structure: the exposure is the stock's lagged weight in each fund portfolio, and the shock is the investor flow hitting that fund. Using lagged rather than contemporaneous weights is essential, since it limits the extent to which the measure captures contemporaneous manager information or discretionary stock selection. The normalized measure divides by lagged stock market equity, and the ranked variable enters the regressions.

Forced-sale pressure is the negative part of flow-induced trading,
\begin{equation}
\ForcedSale_{i,t}=-\min\left\{\frac{\FIT_{i,t}}{ME_{i,t-1}},0\right\}.
\end{equation}
where \(ME_{i,t-1}\) denotes lagged market equity. Scaling by \(ME_{i,t-1}\) converts the dollar selling pressure into a stock-level pressure rate. The inventory proxy, \(\AbsInv\), is a decaying stock of prior forced selling,
\begin{equation}
\AbsInv_{i,t}=\rho\AbsInv_{i,t-1}+\ForcedSale_{i,t},
\label{eq:absinv_empirical}
\end{equation}
with \(\rho=0.85\) in the baseline; the robustness section reports alternative decay parameters. The decaying sum reflects that forced selling need not be absorbed at once by long-horizon natural investors, so that the remaining exposure is carried by marginal risk absorbers and requires compensation. Throughout the empirical section, \(\ForcedSale_{i,t}\) and \(\AbsInv_{i,t}\) are best read as mutual-fund-induced absorption-pressure proxies rather than as complete measures of the total residual inventory in equation \eqref{eq:theta}.

The raw magnitudes are economically meaningful even before ranking. In the full sample, the 90th percentile of flow-induced forced selling is about 24 basis points of market equity and the 95th percentile is about 34 basis points. In the no-microcap sample, the corresponding values are about 29 and 39 basis points. These numbers are averages over monthly stock observations; individual stress episodes can be much larger, especially relative to trading volume.

\begin{figure}[t]
\centering
\resizebox{0.98\textwidth}{!}{
\begin{tikzpicture}[
    box/.style={
        draw=black!55,
        line width=0.4pt,
        fill=black!3,
        text width=2.35cm,
        minimum height=0.88cm,
        align=center,
        font=\footnotesize
    },
    arrow/.style={-{Latex[length=2mm,width=1.4mm]}, draw=black!65, line width=0.45pt}
]
\node[box] (hold) at (0,0.75) {Predetermined\\fund holdings\\\(w_{f,i,t-1}\)};
\node[box] (flow) at (0,-0.75) {Investor flow\\in month \(t\)\\\(Flow_{f,t}\)};
\node[box] (fit) at (3.45,0) {Stock-level\\flow-induced\\trading\\\(\FIT_{i,t}\)};
\node[box] (sale) at (6.45,0) {Forced-sale\\pressure\\\(\ForcedSale_{i,t}\)};
\node[box] (inv) at (9.45,0) {Decaying\\inventory proxy\\\(\AbsInv_{i,t}\)};
\node[box] (ret) at (9.45,-1.65) {Future returns\\\(R_{i,t+1:t+h}\)};

\draw[arrow] (hold.east) -- (fit.west);
\draw[arrow] (flow.east) -- (fit.west);
\draw[arrow] (fit.east) -- (sale.west);
\draw[arrow] (sale.east) -- (inv.west);
\draw[arrow] (inv.south) -- (ret.north);
\end{tikzpicture}
}
\caption{Measurement Timeline}
\label{fig:measurement_timeline}
\begin{minipage}{0.92\textwidth}
\footnotesize
The figure summarizes the construction of the stock-level demand-pressure variables. Fund flows in month \(t\) are mapped through predetermined fund holdings. The negative part of stock-level flow-induced trading is forced-sale pressure; a decaying sum of current and past forced sales is the inventory proxy. The reported-holdings version uses only disclosures available at least 45 days before the flow month.
\end{minipage}
\end{figure}
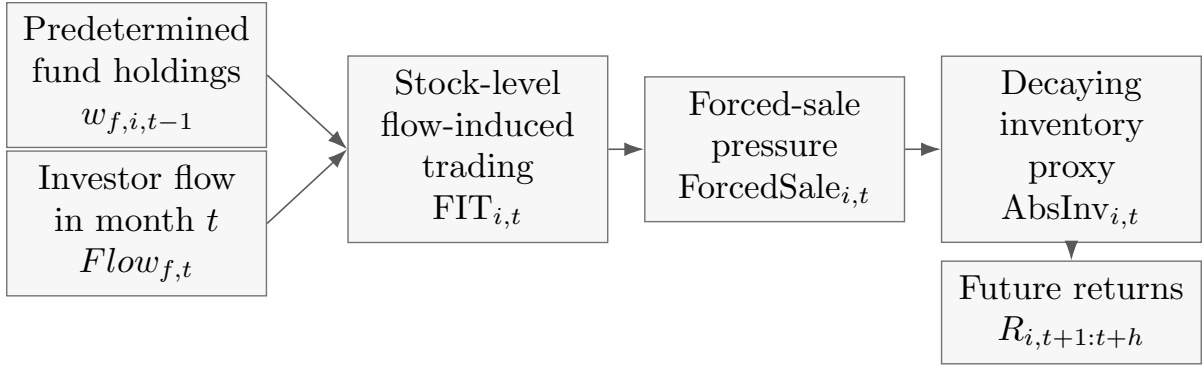

Expected flow pressure is constructed by forecasting fund flows from lagged fund flows, lagged fund returns, fund size, and calendar-month effects, and then mapping predicted flows through current holdings. The reported tests use an expanding-window prediction: for each month, the flow model is estimated only with information available before that month. This measure proxies for the model's expected future residual flow term and asks whether the component of future absorption pressure that is predictable from investor flow dynamics is already reflected in expected returns.

\subsection{Data timing and reporting lags}

Data timing is central to the interpretation. Because mutual fund holdings are reported with a delay and need not be known to market participants at the portfolio date, the baseline flow-holdings measure is best read as an ex post demand-shock measure: it identifies which stocks were exposed to flow-induced pressure given the fund portfolios in place before the flow month. As such, it is well suited to studying price pressure and market clearing, though it is not automatically a real-time trading signal.

We therefore construct a reporting-lag version of the pressure variable. For each fund-month, this version uses only portfolio reports whose report date is at least 45 days before the flow month, and no more than 276 days old. The 45-day lag matches the disclosure delay commonly associated with institutional portfolio reporting. The reported-holdings measure asks whether the return relation survives when the stock-level mapping uses only holdings information that would have been available by the time the flow shock is measured. The results below show that it does. In the no-microcap sample, the reported-holdings forced-sale rank predicts 50, 133, 195, and 321 basis points over one, three, six, and twelve months, with Newey-West t-statistics of 2.83, 3.44, 3.14, and 2.86.

\subsection{Investor base and absorption scarcity}

SEC Form 13F filings measure a stock's natural investor base. Institutional ownership, denoted \(IO\) in some appendix tables, is total 13F shares held divided by shares outstanding; breadth is the log number of institutional owners; and concentration is measured both by the Herfindahl-Hirschman index (HHI) of institutional holdings and by the share of the largest institutional holder. A stock has a thin investor base when ownership and breadth are low and concentration is high, and \(\ThinBase_{i,t}\) is a standardized composite of these components.

Market absorption scarcity is a time-series state variable that combines aggregate Amihud illiquidity, market volatility, and mutual fund redemption stress. High values correspond to periods in which trading capacity is scarce and many final investors are trying to reduce risk at once---states in which the theory predicts inventory pressure to be more expensive.

Table \ref{tab:theory_proxy_map} maps the continuous-time objects to their empirical proxies, recording which measured variable carries each sign restriction into the tests.

\begin{table}[t]
\centering
\begin{threeparttable}
\caption{Theory-to-Measurement Map}
\label{tab:theory_proxy_map}
\footnotesize
\begin{tabularx}{0.96\textwidth}{>{\raggedright\arraybackslash}p{0.22\textwidth}>{\raggedright\arraybackslash}p{0.24\textwidth}>{\raggedright\arraybackslash}p{0.20\textwidth}>{\raggedright\arraybackslash}X}
\toprule
Theory object & Empirical proxy & Predicted role & Main evidence \\
\midrule
\(\dd\theta_t^{MF}\) & \(\ForcedSale_{i,t}\) & Demand shock & Price pressure and future returns \\
\(\theta_t^{MF}\) & \(\AbsInv_{i,t}\) & Inventory state & Future return slopes and portfolios \\
\(\E_t[\dd\theta_t^{MF}]\) & Expected sale pressure & Predictable flow & Out-of-sample tests \\
\(\Phi_t\) and trading capacity & Amihud, turnover, scarcity & Trading cost & Illiquidity and scarcity interactions \\
\(K_t,\ell_t\), natural demand & Scarcity, \(\ThinBase_{i,t}\) & Higher marginal absorption cost & State dependence and mechanism sorts \\
\bottomrule
\end{tabularx}
\begin{tablenotes}[flushleft]
\footnotesize
\item The empirical variables measure the mutual-fund-induced component of residual absorption pressure. They are proxies for testable sign restrictions, not estimates of the primitive cost parameters.
\end{tablenotes}
\end{threeparttable}
\end{table}

\subsection{Regression design}

The main Fama-MacBeth specification is
\begin{equation}
R_{i,t+1:t+h}-R_{f,t+1:t+h}=a_t+b_t\AbsInv_{i,t}+\Gamma_t'Controls_{i,t}+\varepsilon_{i,t+h},
\end{equation}
with Newey-West standard errors applied to the time series of monthly slopes. Parallel specifications use \(\ForcedSale_{i,t}\) and expected future sale pressure. State dependence is tested by allowing slopes to differ between high- and low-scarcity months and by interacting absorption pressure with \(\Scarcity_t\), \(\ThinBase_{i,t}\), and stock-level illiquidity.

Portfolio tests sort stocks into absorption portfolios each month. The central portfolio is high-minus-low \(\AbsInv\). The analysis reports equal-weighted and value-weighted returns and standard factor alphas. The equal-weighted results are important because the model predicts stronger effects where trading capacity is limited. The value-weighted tests provide a conservative check that the results are not entirely confined to the smallest stocks.

\begin{table}[H]
\centering
\begin{threeparttable}
\caption{Sample Coverage and Descriptive Statistics}
\label{tab:descriptive}
\small
\begin{tabular}{lrr}
\toprule
Sample or variable & Mean / count & Standard deviation \\
\midrule
Full sample stock-months & 1,022,728 & \\
Full sample PERMNOs & 10,338 & \\
No-microcap stock-months & 490,162 & \\
No-microcap PERMNOs & 5,559 & \\
Monthly return, full sample (bps) & 96.3 & 1,795.4 \\
Monthly return, no-microcap (bps) & 91.8 & 1,397.2 \\
Forced-sale rank, no-microcap & 0.50 & 0.29 \\
\(\AbsInv\) rank, no-microcap & 0.50 & 0.29 \\
Institutional ownership, no-microcap & 0.61 & 0.31 \\
Breadth rank, no-microcap & 0.50 & 0.29 \\
Amihud rank, no-microcap & 0.50 & 0.29 \\
\bottomrule
\end{tabular}
\begin{tablenotes}[flushleft]
\footnotesize
\item The table summarizes the monthly stock sample used in the empirical analysis. Rank variables are computed within month. Returns are in basis points. Detailed variable definitions are in Appendix Table \ref{tab:variable_definitions}.
\end{tablenotes}
\end{threeparttable}
\end{table}

\section{Main Empirical Results}
\label{sec:results}

\subsection{From fund flows to actual selling}

The return tests begin with a measurement check: if \(\FIT_{i,t}\) is a useful demand-pressure measure, fund flows mapped through lagged holdings should predict actual changes in mutual fund ownership, and Table \ref{tab:first_stage_realtime} confirms that they do. The manager-stock test uses monthly fund-portfolio-by-stock observations, absorbs fund and stock fixed effects within each report month, and asks whether lagged portfolio weights multiplied by investor flows predict actual changes in fund holdings. Because the regression is run report month by report month, the stock fixed effects are stock-by-report-month effects, so the coefficient is identified by comparing funds that hold the same stock in the same month but experience different investor flows. The average monthly slope is 1.07, with a Newey-West t-statistic of 30.24 across 168 report months and 20.5 million manager-stock observations.

The same pattern appears after aggregating to the stock-quarter level. Net flow-induced trading predicts increases in aggregate mutual fund holdings, while sale pressure predicts decreases. A move from the bottom to the top of the quarterly sale-pressure rank is associated with a 114 basis point decline in mutual fund ownership relative to market capitalization, with a t-statistic of -4.48. By contrast, the gross buy-pressure rank does not predict an aggregate mutual fund ownership increase. This distinction matters for the interpretation below: the sell-side measure is validated as actual selling by open-end funds, while the gross buy measure is a broader flow-exposure variable rather than a clean purchase shock. The first stage therefore supports the paper's selling-pressure design without forcing a symmetric buy-side story.

Table \ref{tab:first_stage_realtime} also reports a reporting-lag version of the future-return test. The pressure measure uses only holdings reports at least 45 days old when flow pressure is measured. The no-microcap coefficients remain positive and statistically significant at all horizons, so the return relation is not an artifact of using portfolio disclosures that postdate the shock. The economic interpretation remains one of market clearing rather than of an immediately implementable trading signal.

\begin{table}[t]
\centering
\begin{threeparttable}
\caption{First-Stage and Reporting-Lag Diagnostics}
\label{tab:first_stage_realtime}
\small
\begin{tabular}{lrrrr}
\toprule
Test & 1m / coefficient & 3m & 6m & 12m \\
\midrule
\multicolumn{5}{l}{Panel A: Actual-selling first stages} \\
Manager-stock two-way fixed effects & 1.07 &  &  &  \\
 & (30.24) &  &  &  \\
Net FIT rank \(\rightarrow \Delta\) MF ownership & 52.62 &  &  &  \\
 & (2.40) &  &  &  \\
Sale pressure rank \(\rightarrow \Delta\) MF ownership & -114.24 &  &  &  \\
 & (-4.48) &  &  &  \\
Buy pressure rank \(\rightarrow \Delta\) MF ownership & -27.23 &  &  &  \\
 & (-1.39) &  &  &  \\
\multicolumn{5}{l}{Panel B: Reported-holdings 45-day lag, no-microcap future returns} \\
Reported ForcedSale rank & 50.36 & 132.97 & 195.29 & 321.06 \\
 & (2.83) & (3.44) & (3.14) & (2.86) \\
\bottomrule
\end{tabular}
\begin{tablenotes}[flushleft]
\footnotesize
\item Panel A reports first-stage tests. MF denotes mutual fund. The manager-stock row uses monthly fund-portfolio-by-stock observations with fund and stock fixed effects within report month; the slope is the coefficient on lagged stock weight multiplied by fund flow. The stock-quarter rows report cross-sectional slopes in basis points, where the dependent variable is the change in aggregate mutual fund holdings scaled by market equity. The buy-pressure row is included to show that the gross buy measure is not a clean aggregate purchase shock. Panel B reports Fama-MacBeth slopes using only holding reports at least 45 days old when flow pressure is measured. T-statistics are in parentheses.
\end{tablenotes}
\end{threeparttable}
\end{table}

\subsection{Forced selling, residual inventory, and future returns}

Table \ref{tab:main_fmb} reports the main Fama-MacBeth results, with coefficients expressed in basis points of future cumulative return per one-unit monthly rank increase in the absorption measure. Forced-sale pressure strongly predicts subsequent returns: in the no-microcap sample the one-month slope is 65 basis points, with a t-statistic of 3.32, and it rises to 160 basis points over three months, 217 over six months, and 342 over twelve. The full sample produces similar magnitudes.

\(\AbsInv\) predicts returns as well, with the no-microcap rank delivering 46 basis points over one month, 135 over three months, and 237 over six. The twelve-month coefficient remains economically large but is less precisely estimated. This horizon pattern fits the interpretation that absorption premia are strongest at short and intermediate horizons, while forced inventory is still being carried by marginal capital.

\begin{table}[t]
\centering
\begin{threeparttable}
\caption{Absorption Pressure and Future Returns}
\label{tab:main_fmb}
\small
\begin{tabular}{llrrrr}
\toprule
Measure & Sample & 1m & 3m & 6m & 12m \\
\midrule
ForcedSale rank & All & 73.21 & 162.52 & 225.50 & 344.93 \\
 &  & (3.33) & (3.33) & (2.93) & (2.35) \\
ForcedSale rank & No microcap & 65.43 & 160.11 & 216.50 & 342.33 \\
 &  & (3.32) & (3.97) & (3.54) & (3.07) \\
\(\AbsInv\) rank & All & 68.37 & 174.66 & 273.82 & 337.34 \\
 &  & (3.05) & (3.09) & (2.79) & (1.56) \\
\(\AbsInv\) rank & No microcap & 45.81 & 135.39 & 237.07 & 308.55 \\
 &  & (2.52) & (2.65) & (2.63) & (1.60) \\
\bottomrule
\end{tabular}
\begin{tablenotes}[flushleft]
\footnotesize
\item The table reports monthly Fama-MacBeth slopes. Dependent variables are cumulative future returns over the horizon shown. Coefficients are in basis points. Newey-West t-statistics are in parentheses. Controls include size, book-to-market, momentum, short-term reversal, beta, idiosyncratic volatility, turnover, Amihud illiquidity, institutional ownership, breadth, and concentration.
\end{tablenotes}
\end{threeparttable}
\end{table}

\begin{figure}[t]
\centering
\includegraphics[width=0.76\textwidth]{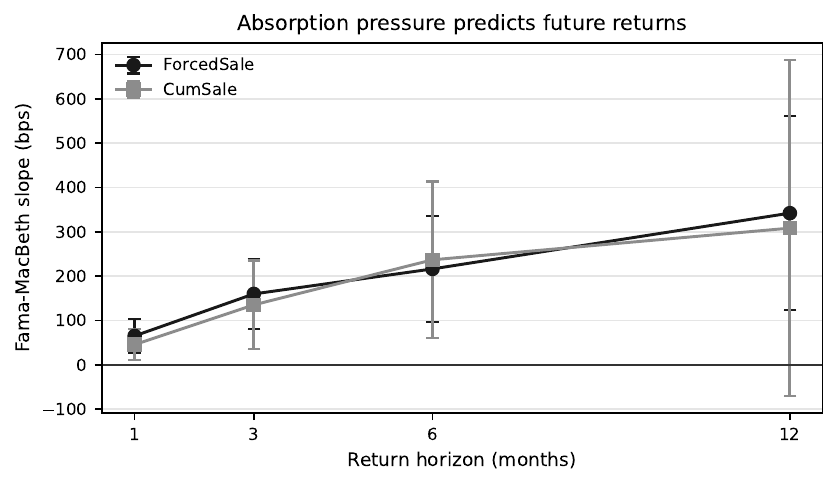}
\caption{Absorption Pressure and Future Return Slopes}
\label{fig:main_coefficients}
\begin{minipage}{0.86\textwidth}
\footnotesize
The figure plots no-microcap Fama-MacBeth slopes from Table \ref{tab:main_fmb}. Coefficients are basis points of cumulative future return for a bottom-to-top move in the monthly rank. Error bars are approximate 95 percent confidence intervals based on Newey-West t-statistics.
\end{minipage}
\end{figure}

Figure \ref{fig:main_coefficients} plots the no-microcap forced-sale and \(\AbsInv\) slopes with confidence intervals. The figure is included to make the horizon pattern visually transparent: the relation is strongest and most precisely estimated over one to six months, while longer horizons remain positive but are more difficult to interpret as pure absorption compensation.

A natural concern is that forced-sale stocks cluster in particular industries. Table \ref{tab:industry_fe} adds industry fixed effects to the cross-sectional regressions, and the short-horizon relation remains positive and statistically meaningful. The coefficient on \(\AbsInv\) in the no-microcap sample is 35 basis points over one month and 97 basis points over three months. The relation weakens at longer horizons once industry effects are absorbed, suggesting that some medium-horizon variation reflects broader sectoral demand and performance cycles, while the short-horizon premium survives industry effects.

\begin{table}[t]
\centering
\begin{threeparttable}
\caption{Cumulative Absorption Pressure with Industry Fixed Effects}
\label{tab:industry_fe}
\small
\begin{tabular*}{0.62\textwidth}{@{\extracolsep{\fill}}lrrrr}
\toprule
Sample & 1m & 3m & 6m & 12m \\
\midrule
All & 47.31 & 107.39 & 143.80 & 103.09 \\
 & (2.31) & (2.04) & (1.56) & (0.52) \\
No microcap & 34.50 & 97.19 & 158.48 & 141.48 \\
 & (2.03) & (1.96) & (1.83) & (0.75) \\
\bottomrule
\end{tabular*}
\begin{tablenotes}[flushleft]
\footnotesize
\item The table reports Fama-MacBeth slopes for \(\AbsInv\) after adding industry fixed effects to each monthly cross section. Coefficients are in basis points and t-statistics are in parentheses.
\end{tablenotes}
\end{threeparttable}
\end{table}

\subsection{State dependence}

The model predicts that the same residual inventory requires greater compensation when absorption capacity is scarce. Table \ref{tab:state_dependence} splits the sample into low- and high-scarcity months. In the full sample, the one-month \(\AbsInv\) slope is 43 basis points in low-scarcity months and 93 basis points in high-scarcity months. The three-month slope rises from 118 to 231 basis points. Similar short-horizon state dependence appears in the no-microcap sample, though the high-scarcity estimates become less precise at longer horizons.

A generic flow-pressure story would predict a positive average reversal, but not that the slope nearly doubles in high-scarcity months. That state dependence is what links the return premium to the cost of taking the other side.

\begin{figure}[t]
\centering
\includegraphics[width=0.76\textwidth]{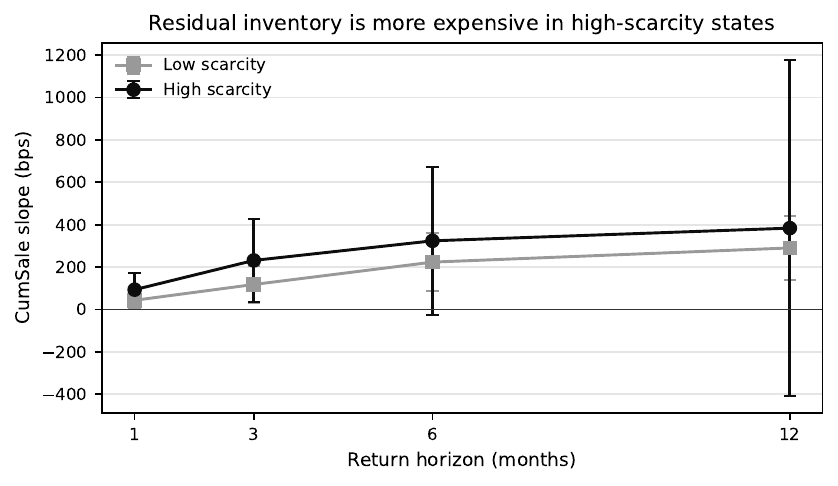}
\caption{State Dependence in Absorption Slopes}
\label{fig:state_dependence}
\begin{minipage}{0.86\textwidth}
\footnotesize
The figure plots full-sample Fama-MacBeth slopes for \(\AbsInv\) in low- and high-scarcity months. Coefficients are in basis points. Error bars are approximate 95 percent confidence intervals based on Newey-West t-statistics.
\end{minipage}
\end{figure}

Figure \ref{fig:state_dependence} shows the same point graphically for the full sample. The visual comparison is useful because state dependence is a key mechanism rather than a purely statistical robustness check. The high-scarcity slope is larger at each reported horizon, with the cleanest separation at short horizons.

Appendix Table \ref{tab:scarcity_decomposition} decomposes the scarcity proxy into market illiquidity, volatility, financial conditions, and mutual fund redemption stress. The individual continuous interactions are mixed, while the nonflow-scarcity coefficient remains positive when redemption stress is removed from the composite. The main text therefore works with the high-versus-low composite state as the cleaner measure of broad absorption-capacity scarcity.

This composite is also the right object given the model. The capital-cost term is nonlinear, so the marginal cost of absorption need not move linearly with every single proxy. Appendix Table \ref{tab:nonparametric_scarcity} repeats the exercise with top- and bottom-tercile state splits. The nonparametric splits show stronger \(\AbsInv\) slopes in high market-illiquidity states, high financial-stress states, high redemption-stress states, and high nonflow-scarcity states at the horizons where the absorption mechanism is sharpest. The nonlinearity of the capital-cost term is precisely why one expects costly-absorption states, viewed jointly, rather than every component in isolation, to carry larger prices of residual supply.

The same logic appears in stock-level capacity measures. Appendix Table \ref{tab:double_sorts} shows that high-minus-low absorption returns are much larger among stocks with high \(\ThinBase\), high illiquidity, or high market scarcity. These are the parts of the cross section where the model predicts that the same residual supply should be most expensive to place.

\begin{table}[t]
\centering
\begin{threeparttable}
\caption{State Dependence: Cumulative Absorption Pressure}
\label{tab:state_dependence}
\small
\begin{tabular}{llrrrr}
\toprule
Sample & State & 1m & 3m & 6m & 12m \\
\midrule
All & Low scarcity & 43.31 & 117.89 & 223.40 & 290.43 \\
 &  & (2.34) & (2.60) & (3.23) & (3.76) \\
All & High scarcity & 93.44 & 231.43 & 323.85 & 383.88 \\
 &  & (2.28) & (2.33) & (1.81) & (0.95) \\
No microcap & Low scarcity & 29.79 & 105.42 & 242.63 & 340.05 \\
 &  & (1.98) & (2.45) & (3.27) & (3.14) \\
No microcap & High scarcity & 61.83 & 165.37 & 231.56 & 277.29 \\
 &  & (1.91) & (1.93) & (1.49) & (0.79) \\
\bottomrule
\end{tabular}
\begin{tablenotes}[flushleft]
\footnotesize
\item The table reports Fama-MacBeth slopes for \(\AbsInv\) separately in low- and high-scarcity months. Scarcity is a composite of market illiquidity, volatility, and mutual fund redemption stress. Coefficients are in basis points.
\end{tablenotes}
\end{threeparttable}
\end{table}

\subsection{Trading capacity conditional on fund ownership}

Size is not the cleanest way to locate absorption costs. Very small stocks can be noisy and may have little mutual fund ownership, while the largest stocks have deep trading markets and broad investor bases. The model instead points to a more specific margin: stocks that are exposed to fund-flow selling but have limited trading capacity. Table \ref{tab:capacity_split} implements this idea by restricting the no-microcap sample to stocks with above-median lagged mutual fund exposure and then splitting them by dollar trading volume.

The pattern is sharp. Among fund-owned stocks with high dollar volume, \(\AbsInv\) slopes are small and statistically weak. Among fund-owned stocks with low dollar volume, the slopes are 97, 375, and 730 basis points over one, three, and six months, with t-statistics of 2.06, 3.79, and 5.14. ForcedSale shows the same ranking, especially at three- and six-month horizons. This table is more informative than a simple size split because it isolates the securities-trading margin in the theory: the premium is largest where funds can create measurable residual supply and where trading capacity is too limited to clear that supply cheaply.

\begin{table}[t]
\centering
\begin{threeparttable}
\caption{Capacity Splits Conditional on Mutual Fund Exposure}
\label{tab:capacity_split}
\small
\begin{tabular}{llrrr}
\toprule
Measure & Capacity state & 1m & 3m & 6m \\
\midrule
ForcedSale & Fund-owned, high dollar volume & 24.11 & 73.79 & 81.14 \\
 &  & (0.85) & (1.06) & (0.79) \\
ForcedSale & Fund-owned, low dollar volume & 50.95 & 268.33 & 507.57 \\
 &  & (1.30) & (3.18) & (3.86) \\
\(\AbsInv\) & Fund-owned, high dollar volume & 4.54 & 39.03 & 91.77 \\
 &  & (0.22) & (0.74) & (0.93) \\
\(\AbsInv\) & Fund-owned, low dollar volume & 96.60 & 375.46 & 729.63 \\
 &  & (2.06) & (3.79) & (5.14) \\
\bottomrule
\end{tabular}
\begin{tablenotes}[flushleft]
\footnotesize
\item No-microcap Fama-MacBeth slopes. The sample is restricted to stock-months with above-median lagged mutual fund exposure, measured by the monthly rank of the sum of lagged mutual fund portfolio weights. High and low dollar-volume states are the top and bottom terciles of lagged twelve-month dollar-volume ranks. Coefficients are in basis points and t-statistics are in parentheses.
\end{tablenotes}
\end{threeparttable}
\end{table}

\subsection{Portfolio and implementation diagnostics}

Portfolio sorts provide a nonparametric view of the return pattern. Each month, stocks are sorted into absorption portfolios using \(\AbsInv\). Table \ref{tab:portfolio_sorts} reports high-minus-low returns. In the no-microcap sample, equal-weighted high-minus-low returns are 44 basis points over one month, 120 basis points over three months, 218 basis points over six months, and 334 basis points over twelve months. The high-scarcity equal-weighted spreads are larger at every horizon. Value-weighted spreads are smaller and less precise, which is consistent with the mechanism being strongest among stocks with limited trading capacity and thinner investor bases.

\begin{table}[t]
\centering
\begin{threeparttable}
\caption{Portfolio Sorts on Cumulative Absorption Pressure}
\label{tab:portfolio_sorts}
\small
\begin{tabular}{llrrrr}
\toprule
Sample & Portfolio & 1m & 3m & 6m & 12m \\
\midrule
No microcap & Equal-weighted H-L & 44.09 & 120.20 & 218.41 & 333.79 \\
 &  & (2.85) & (2.75) & (2.57) & (1.99) \\
No microcap & Value-weighted H-L & 20.92 & 46.17 & 74.92 & 121.80 \\
 &  & (1.29) & (1.06) & (0.96) & (0.96) \\
No microcap & Equal-weighted H-L, high scarcity & 76.21 & 195.41 & 326.79 & 524.75 \\
 &  & (2.64) & (2.89) & (2.46) & (1.89) \\
No microcap & Value-weighted H-L, high scarcity & 28.15 & 68.35 & 84.53 & 112.51 \\
 &  & (1.10) & (1.03) & (0.77) & (0.75) \\
\bottomrule
\end{tabular}
\begin{tablenotes}[flushleft]
\footnotesize
\item Stocks are sorted each month by \(\AbsInv\). The table reports high-minus-low portfolio returns in basis points with Newey-West t-statistics in parentheses.
\end{tablenotes}
\end{threeparttable}
\end{table}

Table \ref{tab:factor_alphas} reports standard factor alphas for the absorption portfolio. The no-microcap equal-weighted portfolio earns 31 basis points per month under CAPM, 35 under the Fama-French three-factor model, 40 under the Carhart four-factor model, and 24 under the Fama-French five-factor-plus-momentum model, while the value-weighted alphas are small. The pattern is informative: absorption pressure is not spanned by standard factors, yet the premium is not a large-cap aggregate factor either. It appears where the market-clearing task is most costly.

Appendix Tables \ref{tab:size_terciles} and \ref{tab:turnover_costs} further discipline the portfolio interpretation. Size alone is an imperfect proxy for absorption capacity: the cumulative-inventory relation is strongest in the middle of the no-microcap size distribution at intermediate horizons and weak among the largest stocks. The equal-weighted absorption portfolio also holds persistent positions, with average one-way long-short turnover of 17.2 percent per month. The gross spread is 47.9 basis points per month and remains 44.0, 39.7, and 31.1 basis points after assumed one-way costs of 25, 50, and 100 basis points. Read as a way to locate the premium rather than as a scalable arbitrage, the portfolio shows that absorption compensation survives conservative trading-cost adjustments precisely where the model predicts absorption to be costly.

\begin{table}[H]
\centering
\begin{threeparttable}
\caption{Factor Alphas of the Absorption Portfolio}
\label{tab:factor_alphas}
\small
\begin{tabular}{llrrrr}
\toprule
Sample & Weight & CAPM & FF3 & Carhart & FF5+UMD \\
\midrule
No microcap & Equal-weighted & 31.28 & 34.72 & 40.21 & 24.46 \\
 &  & (1.79) & (2.19) & (2.51) & (1.83) \\
No microcap & Value-weighted & -8.18 & 1.78 & 5.45 & 9.81 \\
 &  & (-0.51) & (0.14) & (0.43) & (0.83) \\
\bottomrule
\end{tabular}
\begin{tablenotes}[flushleft]
\footnotesize
\item The table reports monthly alphas in basis points for high-minus-low absorption portfolios. T-statistics are in parentheses. FF3 is the Fama-French three-factor model, Carhart adds momentum, and FF5+UMD uses the Fama-French five factors plus momentum.
\end{tablenotes}
\end{threeparttable}
\end{table}

\section{Mechanisms and Identification}
\label{sec:mechanisms}

\subsection{Contemporaneous price pressure and reversal}

The first mechanism test asks whether flow-induced trading moves prices in the direction that demand pressure predicts. Table \ref{tab:price_pressure} shows that net flow-induced trading is positively associated with contemporaneous returns, while forced-sale pressure is strongly negatively associated with them. In the no-microcap sample, the net flow-induced-trading coefficient is 30 basis points and the forced-sale coefficient is \(-62\) basis points, and the price impact is larger in high-scarcity months.

\begin{table}[H]
\centering
\begin{threeparttable}
\caption{Contemporaneous Price Pressure}
\label{tab:price_pressure}
\small
\begin{tabular}{llrr}
\toprule
Sample & State & Net FIT rank & ForcedSale rank \\
\midrule
All & Full period & 42.28 & -79.25 \\
 &  & (4.16) & (-7.51) \\
All & Low scarcity & 26.19 & -58.34 \\
 &  & (2.43) & (-6.67) \\
All & High scarcity & 58.49 & -100.32 \\
 &  & (3.52) & (-5.07) \\
No microcap & Full period & 29.76 & -61.64 \\
 &  & (3.74) & (-6.20) \\
No microcap & Low scarcity & 23.69 & -45.99 \\
 &  & (2.99) & (-6.68) \\
No microcap & High scarcity & 35.88 & -77.41 \\
 &  & (2.56) & (-4.05) \\
\bottomrule
\end{tabular}
\begin{tablenotes}[flushleft]
\footnotesize
\item The dependent variable is the contemporaneous monthly stock return. Net FIT is signed flow-induced trading; ForcedSale is its negative part after scaling by market equity. Coefficients are basis points per rank unit. T-statistics are in parentheses.
\end{tablenotes}
\end{threeparttable}
\end{table}

Together with the first-stage evidence in Table \ref{tab:first_stage_realtime}, the contemporaneous return evidence completes the first two links of the mechanism: fund flows predict selling by funds, and that selling pressure moves prices in the direction required by imperfectly elastic demand.

\begin{figure}[!htbp]
\centering
\includegraphics[width=0.92\textwidth]{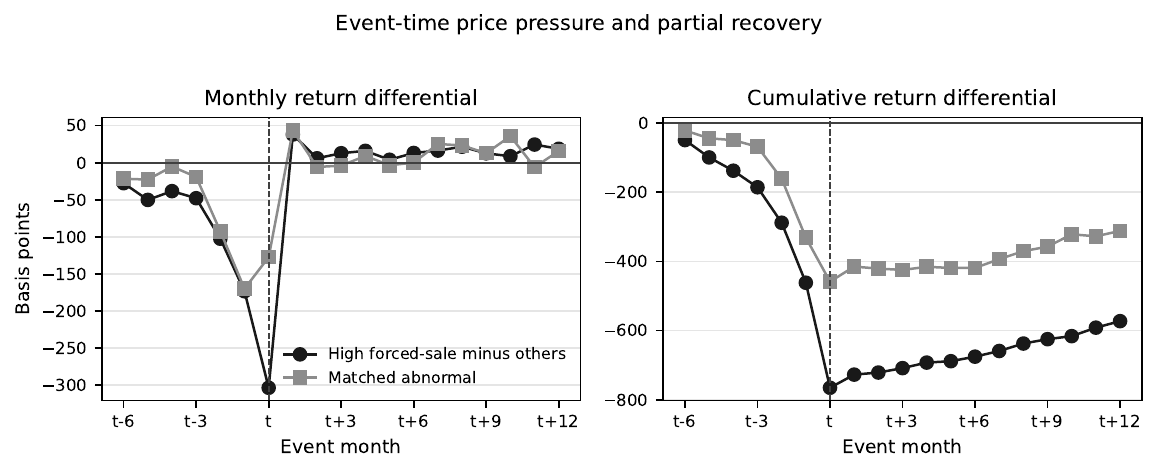}
\caption{Event-Time Price Pressure and Partial Recovery}
\label{fig:event_time}
\begin{minipage}{0.88\textwidth}
\footnotesize
The figure reports event-time returns in basis points around high forced-sale episodes in the no-microcap sample. The black line compares high forced-sale stocks with other stocks. The gray line compares high forced-sale stocks with matched controls formed within month, industry group, size, book-to-market, illiquidity, momentum, and short-term reversal bins. Month 0 is the forced-sale month.
\end{minipage}
\end{figure}

The event-time pattern is consistent with price pressure followed by partial recovery. Figure \ref{fig:event_time} reports average returns for high forced-sale stocks relative to other stocks in the no-microcap sample, together with the corresponding matched abnormal return path. Returns are poor in the months leading up to the event and fall sharply in the event month, and subsequent monthly differentials are mostly positive. This pattern cautions against treating forced-sale shocks as randomly assigned, but it is exactly what one expects when poor prior performance induces redemptions, redemptions create additional selling pressure, and risk absorbers require future compensation to hold the residual supply.

We further repeat the event analysis in a matched sample. High forced-sale stocks are matched within month, industry group, size, book-to-market, illiquidity, momentum, and short-term reversal bins. The matched event path still shows negative returns before the event and a large event-month decline. The matched abnormal return is -127 basis points in month 0 with a t-statistic of -12.12, followed by a positive 44 basis points in month 1 with a t-statistic of 2.00. Later post-event months are positive but uneven. The matched evidence reinforces both parts of the interpretation: forced-sale episodes contain strong price pressure and partial recovery, with the pretrend made explicit rather than assumed away.

\subsection{Distinguishing risk absorption from generic reversal}

The main alternative explanation is ordinary reversal. Forced-sale stocks often have poor prior returns, and any return recovery could be a loser effect rather than compensation for absorption. The tests in this section ask whether that interpretation is sufficient. It is not enough for the paper that forced-sale stocks rebound on average; the rebound should remain after pretrend controls and should be strongest when absorption is costly.

Table \ref{tab:pretrend_controls} includes controls for past stock returns and other pre-event performance measures. The forced-sale effect remains strong. In the no-microcap sample, forced-sale pressure predicts 57 basis points over one month, 139 basis points over three months, 169 basis points over six months, and 272 basis points over twelve months after these controls. \(\AbsInv\) also remains positive at short and intermediate horizons.

We also residualize forced-sale pressure itself on past returns, size, book-to-market, momentum, volatility, illiquidity, turnover, institutional ownership, breadth, and concentration within each month. The residualized forced-sale rank continues to predict one-month returns in the no-microcap sample, with a coefficient of 34 basis points and a t-statistic of 2.05. The coefficients remain positive at longer horizons but become less precise. The cleanest incremental variation is thus concentrated at short horizons, while longer-horizon returns reflect more of prior performance and fundamentals.

Appendix Table \ref{tab:strict_fe} gives a complementary panel check. The one-month forced-sale coefficient remains about 62 basis points after stock fixed effects and industry-month fixed effects. This panel specification complements the main estimator by absorbing persistent stock characteristics and industry-level monthly shocks in a single regression.

\begin{table}[t]
\centering
\begin{threeparttable}
\caption{Future Returns After Pretrend Controls}
\label{tab:pretrend_controls}
\small
\begin{tabular}{llrrrr}
\toprule
Measure & Sample & 1m & 3m & 6m & 12m \\
\midrule
ForcedSale & All & 63.51 & 131.97 & 168.23 & 242.64 \\
 &  & (2.95) & (2.73) & (2.18) & (1.57) \\
ForcedSale & No microcap & 57.49 & 138.62 & 169.11 & 272.26 \\
 &  & (3.12) & (3.72) & (2.80) & (2.35) \\
\(\AbsInv\) & All & 49.58 & 111.50 & 161.75 & 162.66 \\
 &  & (2.42) & (2.18) & (1.74) & (0.76) \\
\(\AbsInv\) & No microcap & 38.04 & 104.43 & 157.93 & 205.07 \\
 &  & (2.47) & (2.55) & (2.06) & (1.23) \\
\bottomrule
\end{tabular}
\begin{tablenotes}[flushleft]
\footnotesize
\item The table reports Fama-MacBeth slopes after controlling for pre-event return patterns. Coefficients are in basis points.
\end{tablenotes}
\end{threeparttable}
\end{table}

Reverse-causality diagnostics confirm that forced-sale pressure is related to poor past performance. Table \ref{tab:reverse_causality} reports regressions of forced-sale pressure on past returns. The negative coefficients are expected: investors redeem from funds after poor performance, and funds sell their lagged holdings. The important point is not that the flow shock is unrelated to past returns. The important point is that pretrends do not absorb the future-return relation, while the state-dependence, thin-base interactions, actual-selling first stage, and placebo tests point to a demand-pressure channel rather than a generic loser rebound alone.

\begin{table}[t]
\centering
\begin{threeparttable}
\caption{Reverse-Causality Diagnostic}
\label{tab:reverse_causality}
\small
\begin{tabular}{lrrr}
\toprule
Sample & Past 1m return & Past 3m return & Past 12m return \\
\midrule
All & -171.01 & -210.10 & -51.05 \\
 & (-3.41) & (-5.55) & (-2.21) \\
No microcap & -65.46 & -324.08 & -162.89 \\
 & (-1.17) & (-6.48) & (-5.73) \\
\bottomrule
\end{tabular}
\begin{tablenotes}[flushleft]
\footnotesize
\item The dependent variable is forced-sale pressure. Coefficients are basis points per rank unit of past returns. T-statistics are in parentheses.
\end{tablenotes}
\end{threeparttable}
\end{table}

\subsection{Ownership migration and investor-base mechanisms}

The absorption theory predicts that forced-sale stocks should become harder to place with stable natural investors, at least in the short run. Table \ref{tab:ownership_migration} shows that high forced-sale pressure predicts declines in institutional ownership and breadth, and increases in ownership concentration, top-holder share, and the thin-base index. In the no-microcap sample, the four-quarter effect on institutional ownership is -455 basis points, while the effect on the thin-base index is 3,092 basis points in rank units. These patterns support the interpretation that forced selling changes the ownership structure in the direction associated with weaker natural absorption capacity.

Appendix Table \ref{tab:absorber_decomposition} decomposes subsequent 13F changes by manager turnover and manager type. The point estimates show continued reductions by the transparent 13F universe, especially among higher-turnover managers, though they are not statistically sharp, and no single class of high-turnover 13F buyer emerges as the quarter-ahead absorber. The picture is thus clear on one side and open on the other: forced-sale stocks leave open-end funds and become harder to place with broad institutional owners, while quarterly 13F filings do not by themselves name the marginal buyer. Absorption may instead run through investors outside 13F, below reporting thresholds, through internal liquidity provision, or through positions too short-lived to appear in quarter-end filings.

\begin{table}[t]
\centering
\begin{threeparttable}
\caption{13F Ownership Migration After Forced Selling}
\label{tab:ownership_migration}
\small
\begin{tabular}{lrr}
\toprule
Outcome & 1 quarter & 4 quarters \\
\midrule
Institutional ownership & -152.54 & -454.95 \\
 & (-6.91) & (-10.18) \\
Breadth & -134.23 & -290.97 \\
 & (-2.75) & (-2.76) \\
HHI & 27.15 & 87.09 \\
 & (1.96) & (2.54) \\
Top-holder share & 30.42 & 107.34 \\
 & (2.02) & (2.69) \\
ThinBase & 1062.36 & 3092.16 \\
 & (4.61) & (6.04) \\
\bottomrule
\end{tabular}
\begin{tablenotes}[flushleft]
\footnotesize
\item The table reports no-microcap regressions of future changes in 13F ownership measures on forced-sale pressure. Coefficients are in basis points of the outcome rank or scaled ownership measure.
\end{tablenotes}
\end{threeparttable}
\end{table}

\subsection{Expected flow pressure, residual redemptions, and placebo tests}

Table \ref{tab:identification} reports additional demand-pressure tests. The first uses distress-sale pressure from funds in the bottom tail of the flow distribution. The second residualizes fund flows each month on lagged fund flows, lagged fund returns, and lagged fund size, then maps the negative residual flow through lagged holdings. This residual-redemption measure is designed to reduce the concern that stock-level pressure simply reflects fund investors responding to fund performance or to the performance of previously held stocks. The third uses expected sale pressure based on expanding-window, out-of-sample predicted fund flows. All three measures predict future returns in the no-microcap sample. Expected sale pressure also interacts strongly with thin investor bases at short and intermediate horizons. This finding is close to the model's flow term: expected future residual supply is priced more strongly when natural demand is weak.

\begin{table}[t]
\centering
\begin{threeparttable}
\caption{Distress Sales, Residual Redemptions, and Expected Flow Pressure}
\label{tab:identification}
\small
\begin{tabular}{lrrrr}
\toprule
Measure & 1m & 3m & 6m & 12m \\
\midrule
DistressSale & 51.33 & 128.33 & 180.79 & 295.80 \\
 & (3.00) & (3.58) & (3.13) & (2.76) \\
Residual redemption sale & 74.08 & 180.08 & 250.23 & 361.80 \\
 & (3.97) & (4.38) & (3.91) & (3.22) \\
Extreme residual redemption sale & 55.20 & 151.10 & 220.41 & 333.59 \\
 & (3.16) & (3.92) & (3.52) & (3.09) \\
ExpectedSaleNext, out-of-sample & 106.86 & 244.10 & 326.65 & 408.18 \\
 & (2.54) & (3.07) & (3.16) & (2.49) \\
ExpectedSaleNext, out-of-sample $\times$ ThinBase & 137.89 & 413.59 & 629.51 & 829.23 \\
 & (2.34) & (2.54) & (2.21) & (1.69) \\
\bottomrule
\end{tabular}
\begin{tablenotes}[flushleft]
\footnotesize
\item The table reports no-microcap Fama-MacBeth slopes. Residual redemption sale uses the negative part of monthly fund-flow residuals after controlling for lagged fund flows, lagged fund returns, and lagged fund size. ExpectedSaleNext is constructed from fund-flow forecasts estimated with expanding windows that use only prior months. Coefficients are in basis points and t-statistics are in parentheses.
\end{tablenotes}
\end{threeparttable}
\end{table}

The placebo evidence is particularly sharp. The flow-holdings mapping is randomly permuted 1,000 times, preserving the marginal distributions but destroying the actual link between fund flows and stock holdings. Table \ref{tab:placebo} compares true coefficients with the 99th percentile of placebo coefficients. Across horizons and samples, the true coefficient is far above the placebo distribution, with empirical p-values of 0.001. This test directly addresses the concern that the results mechanically arise from common time variation, scaling choices, or the cross-sectional distribution of ranks.

Appendix Table \ref{tab:horse_race} runs a horse race against common flow-literature measures. Expected flow pressure remains positive under out-of-sample forecasts, while saturated specifications that combine forced sales, distress sales, expected sales, cumulative pressure, and several interactions deliver unstable individual coefficients, as one would expect when the pressure variables are highly correlated. The incremental evidence therefore comes from the timing and state dependence of the absorption-pressure relation rather than from any single flow proxy subsuming the others.

Appendix Table \ref{tab:purchase_fundamentals} reports two further diagnostics. The purchase side is not a clean mirror image of the sell side. Net signed flow-induced trading is positively related to contemporaneous returns, but when buy and sell pressure enter as separate gross exposure measures, both are associated with weak event-month returns and both predict positive subsequent returns. This rules out a simple one-sided placebo story and indicates that open-end fund flow exposure carries a broader reallocation component, not only redemption fire sales. The absorption interpretation accordingly rests on the selling side, which is the part most directly validated by actual mutual fund ownership declines, contemporaneous negative price pressure, and stronger compensation in costly-absorption states. The same table adds future changes in profitability and sales growth as look-ahead controls to ask whether the return pattern is simply a future-fundamental improvement story. ForcedSale remains positive at the short and intermediate horizons, especially at three months, even as the inventory proxy weakens.

\begin{table}[t]
\centering
\begin{threeparttable}
\caption{Permutation Placebo Tests}
\label{tab:placebo}
\small
\begin{tabular}{llrrrr}
\toprule
Sample & Statistic & 1m & 3m & 6m & 12m \\
\midrule
All & True coefficient & 87.92 & 155.70 & 180.96 & 208.00 \\
All & Placebo 99th pct. & 15.98 & 27.40 & 45.11 & 69.91 \\
All & Empirical p-value & 0.001 & 0.001 & 0.001 & 0.001 \\
No microcap & True coefficient & 77.26 & 155.42 & 183.07 & 249.94 \\
No microcap & Placebo 99th pct. & 18.83 & 27.89 & 44.53 & 69.01 \\
No microcap & Empirical p-value & 0.001 & 0.001 & 0.001 & 0.001 \\
\bottomrule
\end{tabular}
\begin{tablenotes}[flushleft]
\footnotesize
\item The placebo randomly rematches flows and holdings 1,000 times. Coefficients are basis points. The empirical p-value is the share of placebo coefficients at least as large as the true coefficient.
\end{tablenotes}
\end{threeparttable}
\end{table}

\section{Discussion and Scope}
\label{sec:interpretation}

The evidence should be read as a market-clearing result. Mutual funds create a measurable imbalance, but they are not generally the marginal buyers. The marginal absorbers are investors who take the other side after redemptions or carry the inventory while prices recover. The empirical variables are therefore proxies for the pricing restrictions in equation \eqref{eq:euler_quadratic}, not structural estimates of \(\gamma_t\), \(\Phi_t\), \(\ell_t\), or the full absorber balance sheet. \(\ForcedSale\) measures a mutual-fund-induced residual-supply shock, \(\AbsInv\) measures a persistent stock of that observable pressure, expected sales measure predictable future pressure, and \(\Scarcity\) and \(\ThinBase\) shift the marginal cost of absorption.

This interpretation relies on a joint pattern rather than on a single coefficient. Forced sales depress contemporaneous prices because final investors sell into imperfectly elastic demand. Subsequent returns are positive because investors require compensation to carry the position. The compensation is larger when the imbalance is fresh, when aggregate liquidity or funding conditions are poor, when the stock has a thin investor base, or when trading capacity is low. A conventional reversal can explain some rebound after poor returns; it does not naturally predict that the same flow-induced pressure should command a larger premium exactly where absorption is costly.

The evidence also marks the scope of the analysis. Because the flow-holdings mapping is not a randomized experiment and forced selling correlates with poor prior performance, the design leans on a battery of complementary tests---lagged holdings, reporting-lag diagnostics, actual-selling first stages, pretrend controls, residualized pressure, matched event-time analysis, fixed effects, and placebo rematching---that jointly point to a demand-pressure channel. Quarterly 13F data show that forced-sale stocks become harder to place with broad institutional owners but do not isolate a single class of marginal buyers, since some absorption likely runs outside 13F, below reporting thresholds, through internal liquidity provision, or in positions too short-lived to appear at quarter-end.

The portfolio results are best viewed in the same way. Equal-weighted absorption portfolios earn positive alphas in the no-microcap sample, while value-weighted alphas are much smaller. That pattern does not weaken the mechanism. It locates it. The theory predicts the largest compensation where residual supply is costly to clear, not in the deepest large-cap segment where many investors can absorb shocks cheaply. The alphas are therefore diagnostics of a costly market-clearing service rather than evidence of a broad scalable arbitrage factor.

A more structural treatment would estimate the absorber's opportunity set, balance-sheet capacity, and shadow cost of capital directly. The approach here instead tests the observable restrictions the model implies, using holdings, flows, returns, ownership, and liquidity. These restrictions are sharp enough to answer the question at hand: whether one measurable component of residual supply is priced where the marginal cost of absorbing it should be high.

\section{Conclusion}
\label{sec:conclusion}

This paper studies the price of residual supply. The model shows that when final investors leave supply in the market, limited-capital absorbers must be compensated for inventory risk, funding and balance-sheet costs, and costly adjustment, and the empirical tests use mutual fund flows mapped through predetermined holdings to measure one observable component of that supply.

The evidence is consistent with costly risk absorption. Flow-induced selling predicts actual mutual fund sales, contemporaneous price declines, and positive subsequent returns over the horizon in which inventory is plausibly redistributed. The premium is largest when market-wide absorption capacity is tight, investor bases are thin, and trading capacity is low. What the evidence establishes is that a measurable component of residual supply is priced in exactly the states where the theory predicts absorption to be expensive, even as it stops short of tracing the entire marketwide inventory or naming every marginal buyer.

\appendix

\section{Model Derivation Details}

This appendix expands the derivation of the pricing equation used in the paper. The aim is to show that the empirical pricing equation follows from a standard dynamic optimization problem once market clearing assigns residual supply to investors with convex risk, trading, and capital costs. The derivation uses differentiability and interiority conditions locally around the equilibrium path, which suffice for the Euler equation and the comparative statics used in the empirical tests.

The relevant state vector has two parts. The controlled inventory state in the absorber's problem is \(\theta_t^A\), measured as dollar exposure per unit of absorbing capital after raw quantities have been converted through prices and the absorber-capital scale. The market-clearing residual exposure is \(\theta_t^{req}=s_t-n_t(P_t,X_t,Z_t)\). The HJB characterizes the absorber's local demand schedule for \(\theta_t^A\); equilibrium imposes \(\theta_t^A=\theta_t^{req}\). To keep notation readable, the appendix writes the common equilibrium exposure as \(\theta_t\) after this distinction is stated. The exogenous state is \(\mathcal Y_t=(Z_t,K_t,\Sigma_t,\Phi_t,\Psi_t)\), collecting final-demand states, absorbing capital, covariance, trading capacity, and funding conditions. The exogenous state can include diffusion shocks and jumps. The continuous derivation below applies between jumps. Jump shocks are discussed separately because they generate block absorption costs rather than ordinary trading speeds.

The derivation can be made more explicit by writing the state vector as
\[
\mathcal Y_t=(Z_t,K_t,\Sigma_t,\Phi_t,\Psi_t),
\]
and treating \(\theta_t\) as the endogenous controlled inventory state of the absorber. The derivation is local and price-taking. The absorber takes the expected excess return vector \(\pi\) and the law of the exogenous state \(Y\) as given when solving its control problem; the equilibrium step later asks which \(\pi\) makes the optimal inventory path consistent with market clearing. In the local stationary derivation, \(\pi\) is a fixed local return vector. If expected returns vary with exogenous states, one can enlarge \(Y\) to include the return state or write \(\pi=\pi(Y)\). In either case, \(\pi\) is held fixed with respect to \(\theta\) in the absorber's price-taking problem. This distinction is important because the derivative below is a partial derivative in the absorber's problem, not a total derivative of an equilibrium price function.

Suppose the exogenous states follow a diffusion with generator \(\mathcal{L}^{Y}\). In the baseline appendix derivation, the coefficients of this generator depend on \(y\) but not directly on the individual absorber's controlled inventory \(\theta\). For example,
\[
\mathcal L^Y V(\theta,y)=b_Y(y)^{\T}V_y(\theta,y)
+\frac{1}{2}\operatorname{tr}\!\left(a_Y(y)V_{yy}(\theta,y)\right).
\]
This assumption makes \(Y\) exogenous in the absorber's local problem and lets \(\nabla_\theta(\mathcal L^YV)=\mathcal L^Y(V_\theta)\). If the drift, volatility, or jump intensity of \(Y\) depended directly on \(\theta\), the envelope equation would contain the corresponding derivatives of the generator coefficients. Such terms are outside the baseline local pricing restriction and can be interpreted as additional state-feedback or hedging components.

The risk absorber's value function is \(V(\theta,y)\), where \(\theta\) denotes the absorber's controlled inventory state \(\theta^A\) and \(y\) denotes a realization of \(Y_t=\mathcal Y_t\). If the continuous finite-variation component of inventory satisfies \(\dd \theta_t^A=u_t\dd t\), the dynamic programming equation is
\begin{equation}
\rho V(\theta,y)=\max_u\left\{\theta^{\T}\pi-c(\theta,u,K)+V_\theta(\theta,y)^{\T}u+\mathcal{L}^{Y}V(\theta,y)\right\}.
\label{eq:hjb_appendix}
\end{equation}
The first-order condition is
\begin{equation}
\nabla_u c(\theta,u,K)=V_\theta(\theta,y).
\label{eq:foc_appendix}
\end{equation}
This condition states that the marginal execution cost of changing inventory equals the shadow value of inventory. If trading is cheap, this shadow value is small and the risk absorber rapidly moves toward the desired position. If trading is costly, the shadow value becomes a state variable and expected returns must compensate the absorber for carrying positions away from the frictionless target.

The envelope step behind the next equation is as follows. Define the Hamiltonian
\[
\mathcal H(\theta,u;y,\pi)
=\theta^{\T}\pi-c(\theta,u,K)
+V_\theta(\theta,y)^{\T}u+\mathcal L^YV(\theta,y),
\]
and let \(u^*(\theta,y)\) be an interior maximizer. Differentiating the maximized right-hand side gives
\[
\frac{\partial}{\partial\theta}\mathcal H(\theta,u^*(\theta,y);y,\pi)
+\frac{\partial \mathcal H}{\partial u}(\theta,u^*(\theta,y);y,\pi)
\frac{\partial u^*(\theta,y)}{\partial\theta}.
\]
The second term is zero by the first-order condition \(\partial \mathcal H/\partial u=0\). Hence no \(u_\theta\) term appears in the envelope equation. Holding \(\pi\) fixed in the absorber's price-taking problem, the remaining partial derivatives are
\[
\frac{\partial}{\partial\theta}(\theta^{\T}\pi)=\pi,\qquad
\frac{\partial}{\partial\theta}\{-c(\theta,u,K)\}=-\nabla_\theta c(\theta,u,K),
\]
\[
\frac{\partial}{\partial\theta}\{V_\theta(\theta,y)^{\T}u\}=V_{\theta\theta}(\theta,y)u,\qquad
\frac{\partial}{\partial\theta}\{\mathcal L^YV(\theta,y)\}=\nabla_\theta(\mathcal L^YV)(\theta,y).
\]
The envelope condition is therefore
\begin{equation}
\rho V_\theta
=\pi-\nabla_\theta c(\theta,u,K)+V_{\theta\theta}u+\nabla_\theta(\mathcal{L}^{Y}V).
\label{eq:envelope_raw}
\end{equation}
This is the formula used in the paper. It is not an unconditional identity. If the absorber internalized an equilibrium return function \(\pi=\pi(\theta,y)\), then the first derivative would instead be
\[
\frac{\partial}{\partial\theta}\{\theta^{\T}\pi(\theta,y)\}
=\pi(\theta,y)+[\nabla_\theta\pi(\theta,y)]^{\T}\theta,
\]
and the additional term \([\nabla_\theta\pi]^{\T}\theta\) would enter equation \eqref{eq:envelope_raw}. The paper omits this term because the absorber is price-taking in the control problem. Market clearing then determines the \(\pi\) that supports the required residual inventory.

Define the controlled generator of the full state \((\theta^A,Y)\) by
\begin{equation}
\mathcal A^u f(\theta,y)
=u^{\T}\nabla_\theta f(\theta,y)+\mathcal L^Y f(\theta,y).
\label{eq:appendix_controlled_generator}
\end{equation}
Let \(p(\theta,y)=V_\theta(\theta,y)\). Under the exchangeability condition above,
\[
\mathcal A^u p(\theta,y)
=V_{\theta\theta}u+\mathcal L^Y(V_\theta)
=V_{\theta\theta}u+\nabla_\theta(\mathcal L^YV).
\]
Thus \(\mathcal A^u p\) is the predictable drift of the inventory shadow value along the controlled inventory path. Substituting this expression into \eqref{eq:envelope_raw} and using \eqref{eq:foc_appendix} yields the baseline Euler equation
\begin{equation}
\pi=\nabla_\theta c(\theta,u,K)-\mathcal A^u p(\theta,y)+\rho p(\theta,y),
\qquad p(\theta,y)=\nabla_u c(\theta,u,K).
\label{eq:appendix_euler_generator}
\end{equation}
This is equation \eqref{eq:euler_general}. If one works with a smaller state vector that omits some future demand, capital, or opportunity-state risks, the omitted hedging component can be represented by an additive \(h\) term. Under the complete-state interpretation above, \(h=0\); under the smaller-state empirical representation, \(h\) summarizes incremental hedging demand rather than a second copy of the predictable drift. The proof is intentionally written without imposing a representative-investor SDF at the outset. The SDF representation can be recovered after the risk absorber's marginal value has been characterized.

The unit normalization in the paper is also important. Let \(\bar q^A\) be the residual position in raw share units. The required exposure implied by market clearing, and the controlled exposure used in the dynamic program after imposing \(\theta^A=\theta^{req}\), is
\begin{equation}
\theta=\frac{1}{K^A}\diag(P)\bar q^A,
\end{equation}
where \(K^A\) is the dollar capital of the absorbing sector used to scale positions. If \(\theta\) were measured in shares, \(\nabla_\theta c\) would be a dollar cost per share and would not be directly comparable to an expected return. The empirical and theoretical object is instead a dollar exposure per unit of absorbing capital. Under this normalization, \(\theta^{\T}\pi\) is an instantaneous excess return on the absorber's capital, and \(c(\theta,u,K)\) is a cost rate measured in the same units. The marginal condition therefore prices a unit increase in exposure as a return compensation. This normalization is the reason equation \eqref{eq:euler_quadratic} can be taken to the cross section without an additional price-level scaling term.

The capital-cost term provides the nonlinear comparative static. For
\begin{equation}
\ell(\theta,K)=\frac{1}{2}\theta^{\T}\Psi\theta+\frac{\eta}{2}\left[\max\{0,m^{\T}|\theta|-K\}\right]^2,
\end{equation}
the marginal capital cost on a region with fixed position signs is
\begin{equation}
\nabla_\theta\ell(\theta,K)
=\Psi\theta+\eta\left[\max\{0,m^{\T}|\theta|-K\}\right]\1_{\{m^{\T}|\theta|>K\}}\,m\odot \operatorname{sgn}(\theta),
\label{eq:capital_gradient}
\end{equation}
where \(\odot\) denotes element-by-element multiplication. At \(\theta_i=0\) or \(m^{\T}|\theta|=K\), this expression should be interpreted as a subgradient, or as the limiting gradient of a smooth approximation to \(|\theta|\) and the positive-part operator. Equation \eqref{eq:capital_gradient} shows why the price of absorption can rise sharply in stressed states. When the capital constraint is slack, the marginal cost is locally linear. Once required capital exceeds available capacity, the marginal cost rises with the excess required capital. This term is the theoretical source of the empirical interaction between absorption pressure and scarcity.

To connect the model to an SDF, consider a margin on which the risk absorber can trade without an additional wedge after paying the shadow costs summarized above. In a wealth-based formulation consistent with the reduced-form cost representation, the absorber's marginal value can be represented by an SDF with
\begin{equation}
\frac{\dd M_t^A}{M_t^A}=-r_t\dd t-(\lambda_t^A)^{\T}\dd W_t.
\end{equation}
If the only priced diffusion risk were inventory risk with quadratic cost, the diffusion risk price would contain the term
\begin{equation}
\lambda_t^{inv}=\gamma_t\sigma_t^{\T}\theta_t.
\end{equation}
Capital, funding, and demand-state risks add further components to \(\lambda_t^A\), while trading and funding frictions add wedges on margins where positions cannot be adjusted freely. Thus the market-clearing equation and the no-arbitrage equation are compatible representations of the same local pricing restriction. The no-arbitrage equation prices payoffs using marginal value; the market-clearing equation explains why that marginal value is high in states where residual supply is costly to absorb.

Finally, the stock-flow empirical equation follows from a local Taylor expansion of \eqref{eq:euler_quadratic}. Around a reference state \((\bar{\theta},\bar{u},\bar{\kappa})\), write
\begin{equation}
\pi_i \approx a_i
+A_i(\theta_i-\bar{\theta}_i)
+B_i\E_t[\Delta \theta_i^{res}]
+C_i(u_i-\bar{u}_i)
+G_i(\kappa-\bar{\kappa})
+H_i(\theta_i-\bar{\theta}_i)(\kappa-\bar{\kappa}).
\end{equation}
The empirical variables are ranked proxies for these local states. Forced-sale pressure measures the innovation to residual supply, \(\AbsInv\) measures the inventory state, expected sale pressure measures a predictable future residual-flow component, and the scarcity and thin-base interactions measure cross-sectional and time-series variation in the marginal cost of absorption. These terms are projection components of the expected-return restriction; they are not separate structural primitives added on top of a complete-state Euler equation. The Taylor expansion is not used to estimate a fully structural model; it is used to translate the continuous-time equilibrium condition into testable cross-sectional restrictions.

\subsection{A more explicit equilibrium interpretation}

The dynamic program takes prices and expected returns as given, but the equilibrium interpretation closes the model through market clearing. In raw quantities, final demand \(\bar D(P,X,Z)\) and residual absorber demand \(\bar q^A\) satisfy
\begin{equation}
\bar D(P_t,X_t,Z_t)+\bar q_t^A=\bar S_t.
\end{equation}
The pricing equation is written after converting this raw clearing condition into normalized dollar exposures:
\begin{equation}
n_t(P_t,X_t,Z_t)+\theta_t^{req}=s_t,
\qquad
\theta_t^{req}=\frac{1}{K_t^A}\diag(P_t)\bar q_t^A.
\end{equation}
Given a price vector, the final-demand system determines how much normalized supply is left for risk absorbers. Given a controlled inventory state \(\theta_t^A\), the risk absorber's Euler equation determines the expected return required to hold that exposure. The equilibrium price process is one for which these two conditions are consistent and \(\theta_t^A=\theta_t^{req}\).

Locally, the market-clearing equation can be written as
\begin{equation}
n_{P,t}\dd P_t+\dd\theta_t^{req}=\dd s_t-n_{X,t}\dd X_t-n_{Z,t}\dd Z_t.
\label{eq:local_clearing_appendix}
\end{equation}
When normalized dollar demand is strongly price responsive in the relevant direction, price concessions can induce final investors to absorb shocks and required residual exposure changes little. This condition is not the same as raw share demand being downward sloping. In one dimension, \(n(P)=P\bar D(P)/K^A\) gives \(n_P(P)=(\bar D(P)+P\bar D'(P))/K^A\), so \(n_P<0\) requires dollar demand to be sufficiently elastic. More generally, the relevant local object is the price response of normalized required residual exposure, \(s_{P,t}-n_{P,t}\), or equivalently the symmetric part of the normalized dollar-demand derivative when the supply-scaling term is held fixed. If this response is weak, final investors are effectively inelastic and shocks pass through to \(\theta_t^{req}\). The normalized notation matters here. If the primitive clearing equation is written in shares, the differential of \(\theta_t^{req}=(K_t^A)^{-1}\diag(P_t)\bar q_t^A\) contains terms from price changes and changes in absorber capital in addition to changes in raw quantities. Equation \eqref{eq:local_clearing_appendix} is therefore the clean local expansion of the market-clearing requirement in the normalized exposure space used by the Euler equation, not an expansion in raw share units and not the controlled inventory law \(\dd\theta_t^A=u_t\dd t\). The expected-return equation is state dependent even before considering risk. A given flow shock creates a large expected-return response only when it creates residual exposure that is costly for marginal holders to absorb.

Under these monotonicity conditions, the local comparative-static mapping between residual exposure and required returns is well behaved. The risk absorber's marginal cost is increasing in normalized exposure, and price concessions reduce the normalized residual exposure that must be absorbed. In one dimension, market clearing defines normalized residual exposure \(\theta(P)=s(P)-n(P,X,Z)\), and the absorber requires an expected return \(g(\theta)\) with \(g'(\theta)>0\). The stable sign condition is \(\theta_P(P)=s_P(P)-n_P(P)>0\), so that a lower price reduces \(\theta\), while a larger \(\theta\) raises the required return. This condition can hold because supply is also valued in dollars, but it is not implied by \(\bar D'(P)<0\) alone. In multiple dimensions, the analogous condition is imposed on the relevant symmetric part of the local residual-exposure response together with the positive definiteness of the absorber's marginal-cost matrix.

This is a local stability condition for the price-pressure comparative static, not a global uniqueness theorem. A full uniqueness proof would require specifying cash-flow dynamics, beliefs, and the mapping from prices into future expected returns. The empirical tests use comparative statics around observed market states. The restrictions that matter are the signs of the marginal cost terms and how those signs change with absorption capacity.

\subsection{Capital cost and state-dependent convexity}

The capital-cost example in the main text is useful because it generates the nonlinear state dependence that the empirical section tests. The marginal cost in equation \eqref{eq:capital_gradient} has two parts. The first, \(\Psi\theta\), is a smooth financing or balance-sheet charge. The second appears only when required capital exceeds available capacity:
\begin{equation}
\eta\left[m^{\T}|\theta|-K\right]m\odot \operatorname{sgn}(\theta)
\quad\text{when}\quad m^{\T}|\theta|>K.
\end{equation}
The cross-partial derivative with respect to exposure and capital is negative in the constrained region:
\begin{equation}
\frac{\partial}{\partial K}\nabla_\theta\ell(\theta,K)
=-\eta\,m\odot \operatorname{sgn}(\theta).
\end{equation}
An increase in available capital lowers the marginal required return for positions that consume capital. Equivalently, a decline in capital raises the marginal required return. The positive scarcity-return prediction in the empirical analysis is stated for the forced-sale side, where absorbers are required to hold positive residual exposure. For short residual exposure, the sign of the position-specific marginal cost should be interpreted with the corresponding sign of \(\theta_i\). This comparative static is the formal counterpart of the empirical prediction that forced-sale absorption pressure should be priced more strongly in high-scarcity states.

The same logic applies to trading capacity, with the same sign caveat as in the main text. If \(\Phi_t\) rises, the same trading speed \(u_t\) has a higher shadow cost \(p_t=\Phi_tu_t\) under the quadratic adjustment-cost example. Whether the Euler term \(-\mathcal A^{u_t}p+\rho p_t\) rises depends on the direction of adjustment and on the expected path of the shadow cost. Low market depth or high price impact can raise expected returns when the costly trade is part of taking or carrying positive residual exposure, but dynamic trading costs can also reduce the premium if future demand or liquidation paths lower the costate quickly.

The convexity of \(c(\theta,u,K)\) also gives a clear cross-sectional prediction. Suppose two stocks have the same forced-sale shock but different \(m_i\), the capital weight or financing haircut. The stock with the higher \(m_i\) has a higher marginal capital cost in stressed states. Similarly, if two stocks have the same residual supply but different rows of \(\Sigma_t\), the stock more correlated with the inventory portfolio has the higher inventory premium. These predictions motivate controls for ordinary characteristics while testing whether absorption variables retain explanatory power.

\subsection{SDF representation from the absorber's marginal value}

Let \(W_t^A\) denote the wealth of the representative marginal absorber sector. The value of an additional unit of wealth is \(V_W\). In a wealth-based formulation consistent with the reduced-form cost representation above, the absorber's marginal value can be represented by an SDF proportional to \(V_W\):
\begin{equation}
M_t^A=e^{-\int_0^t r_s\dd s}\frac{V_W(W_t^A,\theta_t,\mathcal Y_t)}{V_W(W_0^A,\theta_0,\mathcal Y_0)}.
\end{equation}
Applying Ito's lemma gives
\begin{equation}
\frac{\dd M_t^A}{M_t^A}=-r_t\dd t-(\lambda_t^A)^{\T}\dd W_t+\text{jump terms}.
\end{equation}
For a traded diffusion payoff on an unconstrained margin, no-arbitrage implies
\begin{equation}
\pi_t=\sigma_t\lambda_t^A.
\end{equation}
This is a compatible representation, not a separate primitive derivation from utility, wealth dynamics, and all budget constraints. The market-clearing derivation identifies components that such an absorber-implied risk price would contain. With quadratic inventory cost, the direct inventory component is \(\gamma_t\sigma_t^{\T}\theta_t\). If capital shocks enter the marginal value of wealth, there is an additional component proportional to the covariance between the asset return and the capital state. If demand shocks affect future investment opportunities, there is a demand hedging component. If trading constraints bind, not every expected return component can be represented as a diffusion risk price; the remaining part is a wedge on the constrained trading margin.

This derivation also fixes the role of the SDF. Expected returns are still tied to marginal value. The model explains why marginal value can load on demand and capital states in the first place. In a frictionless representative-agent model, those states may be irrelevant. In a market-clearing model with limited risk absorption, they matter because they determine how costly it is to clear residual supply.

\subsection{Jump shocks and block absorption costs}

The continuous trading-speed term \(u_t\) is not the right object for a discontinuous flow shock. If normalized final demand jumps because \(Z_t\) changes, residual exposure jumps by
\begin{equation}
\Delta\theta_t^{res}=-n_{Z,t}\Delta Z_t+\Delta s_t+\text{higher-order terms}.
\end{equation}
A block absorption cost can be written as
\begin{equation}
\mathcal{J}_t(\Delta\theta)
=\frac{1}{2}\Delta\theta^{\T}\Omega_t\Delta\theta
+\xi_t^{\T}|\Delta\theta|,
\end{equation}
where \(\Omega_t\) captures block price impact and is positive semidefinite, and \(\xi_t\geq 0\) captures fixed or proportional balance-sheet costs of taking the block. Away from coordinates with \(\Delta\theta_i=0\), the marginal compensation for absorbing a jump is
\begin{equation}
\nabla_{\Delta\theta}\mathcal{J}_t(\Delta\theta)
=\Omega_t\Delta\theta+\xi_t\odot\operatorname{sgn}(\Delta\theta).
\end{equation}
At \(\Delta\theta_i=0\), the absolute-value term is interpreted as a subgradient, or equivalently as the limit of a smooth approximation.
This term is the jump analogue of the continuous trading and inventory costs. It predicts a large contemporaneous price concession when the block arrives and positive subsequent expected returns if the block is held by risk absorbers while being redistributed. The event-time tests in the paper are designed to capture this jump-like component of the theory.

The jump formulation also explains why short-horizon returns are central. A large demand shock can create an immediate price effect even if the long-run cash-flow value of the asset is unchanged. If capital is slow moving, the price does not jump all the way back immediately. Instead, expected returns are temporarily high while the market compensates holders for carrying the inventory. This is the continuous-time version of the empirical price-pressure and reversal pattern.

\subsection{Boundary cases}

The model contains several standard asset-pricing mechanisms as boundary cases. If \(\Phi_t=0\), \(\ell_t=0\), and \(h_t=0\), the pricing equation reduces to
\begin{equation}
\pi_t=\gamma_t\Sigma_t\theta_t.
\end{equation}
If \(\theta_t\) is proportional to the market portfolio, this equation has the form of a continuous-time CAPM. If \(\theta_t\) instead represents the balance sheet of a constrained intermediary sector, the same equation becomes an intermediary asset-pricing relation. The difference is not algebraic; it is the identity of the portfolio that market clearing assigns to the marginal sector.

If inventory costs are small but trading costs are large, expected returns are dominated by
\begin{equation}
-\mathcal A^{u_t}p(\theta_t^A,Y_t)+\rho p_t,
\qquad p_t=\Phi_tu_t
\end{equation}
in the quadratic adjustment-cost example. This case generates slow adjustment. Prices move when demand shocks arrive, and expected returns remain elevated until the costly trading program is completed. If capital costs are the dominant term, expected returns depend most strongly on \(\nabla_\theta\ell_t(\theta_t,K_t)\), and the central prediction is state dependence with respect to \(K_t\). If demand-state hedging dominates, the relevant assets are those that perform poorly exactly when future demand imbalances or capital scarcity become worse.

These boundary cases are not separate theories competing with the paper. They are different regions of the same market-clearing equation. The empirical design tests whether the inventory, flow, and scarcity terms have the signs implied by that equation in U.S. equities.

\section{Variable Definitions}

\begin{table}[H]
\centering
\begin{threeparttable}
\caption{Variable Definitions}
\label{tab:variable_definitions}
\small
\begin{tabularx}{\textwidth}{lX}
\toprule
Variable & Definition \\
\midrule
\(TNA_{f,t}\) & Total net assets of mutual fund \(f\) at the end of month \(t\). \\
\(R_{f,t}\) & Net return of mutual fund \(f\) in month \(t\). \\
\(ME_{i,t}\) & Market equity of stock \(i\), used to scale dollar demand pressure. \\
\(Flow_{f,t}\) & Mutual fund net investor flow rate, computed from total net assets after removing fund returns. \\
\(\FIT_{i,t}\) & Flow-induced trading: the dollar demand pressure obtained by multiplying fund flows by lagged stock portfolio weights and lagged total net assets. \\
\(\ForcedSale_{i,t}\) & Negative part of stock-level flow-induced trading, scaled by lagged market equity. \\
\(\AbsInv_{i,t}\) & Decaying stock of forced-sale pressure with baseline \(\rho=0.85\). \\
ExpectedSaleNext & Stock-level sale pressure implied by expanding-window predicted fund flows and current holdings. \\
Scarcity & Composite of aggregate illiquidity, volatility, and mutual fund redemption stress. \\
ThinBase & Composite of low institutional ownership, low breadth, and high concentration. \\
Amihud & Average daily absolute return divided by dollar volume. \\
IO & Institutional ownership, computed as total 13F shares held divided by shares outstanding. \\
Breadth & Log number of 13F institutional owners. \\
HHI & Herfindahl-Hirschman index of institutional holdings. \\
\bottomrule
\end{tabularx}
\end{threeparttable}
\end{table}

\begingroup
\footnotesize
\setlength{\tabcolsep}{3.6pt}
\renewcommand{\arraystretch}{0.94}
\setlength{\LTpre}{4pt}
\setlength{\LTpost}{6pt}
\begin{longtable}{p{0.13\textwidth}p{0.27\textwidth}p{0.17\textwidth}p{0.33\textwidth}}
\caption{Variable--Equation Index}
\label{tab:variable_equation_index}\\
\toprule
Variable & Meaning & Equation(s) & Role in the model \\
\midrule
\endfirsthead
\multicolumn{4}{l}{\footnotesize Table \ref{tab:variable_equation_index} continued}\\
\toprule
Variable & Meaning & Equation(s) & Role in the model \\
\midrule
\endhead
\midrule
\multicolumn{4}{r}{\footnotesize Continued on next page}\\
\endfoot
\bottomrule
\endlastfoot
\multicolumn{4}{l}{\textit{Panel A. Returns, Risk Prices, and SDF Variables}}\\
\addlinespace[2pt]
\(t,s\) & Time indices & Continuous-time equations & Index current and future points on the equilibrium path. \\
\(i\) & Individual asset index & \eqref{eq:inventory_covariance}, \eqref{eq:inventory_decay} & Converts vector restrictions into asset-level implications. \\
\(N\) & Number of risky assets & Text definition & Determines the dimension of \(R_t\), \(\pi_t\), \(\theta_t\), and related vectors. \\
\(m\) & Number of Brownian shocks & \eqref{eq:return_process}, \eqref{eq:sdf_diffusion}, \eqref{eq:absorber_sdf} & Determines the dimension of \(W_t\) and \(\lambda_t\). \\
\(R_t\) & Vector of risky-asset total returns & \eqref{eq:return_process}, \eqref{eq:inventory_covariance} & Describes the stochastic return process of traded assets. \\
\(\dd R_t\) & Instantaneous change in returns & \eqref{eq:return_process}, \eqref{eq:inventory_covariance} & Separates expected returns from risk shocks. \\
\(\mu_t\) & Conditional expected return vector & \eqref{eq:return_process}; \(\pi_t=\mu_t-r_t\mathbf 1\) & Forms expected excess returns. \\
\(r_t\) & Risk-free rate & \(\pi_t=\mu_t-r_t\mathbf 1\); \eqref{eq:sdf_diffusion}, \eqref{eq:absorber_sdf} & Converts total returns into excess returns. \\
\(\mathbf 1\) & Vector of ones & \(\pi_t=\mu_t-r_t\mathbf 1\) & Subtracts the risk-free rate asset by asset. \\
\(\sigma_t\) & Volatility matrix & \eqref{eq:return_process}, \eqref{eq:inventory_cost}, \eqref{eq:inventory_risk_price} & Determines risk exposures and return covariance. \\
\(W_t\) & Brownian motion & \eqref{eq:return_process}, \eqref{eq:sdf_diffusion}, \eqref{eq:absorber_sdf} & Represents the primitive diffusion shocks. \\
\(\Sigma_t\) & Return covariance matrix, \(\sigma_t\sigma_t^\top\) & \eqref{eq:sigma_process}, \eqref{eq:inventory_cost}, \eqref{eq:euler_quadratic}, \eqref{eq:inventory_covariance} & Determines the cost of inventory risk. \\
\(\pi_t\) & Expected excess return vector & \eqref{eq:pricing_wedge}, \eqref{eq:static_absorption}, \eqref{eq:absorber_objective}, \eqref{eq:euler_general}, \eqref{eq:euler_quadratic}, \eqref{eq:stock_flow} & Main object explained by absorption costs. \\
\(M_t\) & General stochastic discount factor & \eqref{eq:sdf_diffusion} & Provides the no-arbitrage pricing background. \\
\(\lambda_t\) & Brownian market price of risk & \eqref{eq:sdf_diffusion}, \eqref{eq:pricing_wedge} & Represents standard SDF risk prices. \\
\(\omega_t\) & General pricing wedge & \eqref{eq:pricing_wedge} & Captures nonfrictionless terms from trading, funding, or constraints. \\
\(M_t^A\) & Risk absorber's SDF & \eqref{eq:absorber_sdf} & Rewrites the SDF in terms of the absorber's marginal value. \\
\(\lambda_t^A\) & Risk absorber's implied risk price & \eqref{eq:absorber_sdf} & Risk price when the absorber is marginal on an unconstrained margin. \\
\(\lambda_t^{inv}\) & Inventory risk price & \eqref{eq:inventory_risk_price} & Converts \(\gamma_t\sigma_t^\top\theta_t\) into a diffusion risk price. \\
\(\omega_t^{abs}\) & Absorption wedge & \eqref{eq:wedge_decomposition}, \eqref{eq:inventory_risk_price} & Collects wedges from capital, trading, funding, and hedging costs. \\
\addlinespace[3pt]
\multicolumn{4}{l}{\textit{Panel B. Market Clearing and Residual Exposure Variables}}\\
\addlinespace[2pt]
\(P_t\) & Price vector & \eqref{eq:raw_demand}, \eqref{eq:raw_residual}, \eqref{eq:normalized_supply_demand}, \eqref{eq:theta_expansion} & Affects final demand and converts shares into dollar exposure. \\
\(X_t\) & Asset characteristics and cash-flow states & \eqref{eq:raw_demand}, \eqref{eq:raw_residual}, \eqref{eq:normalized_supply_demand}, \eqref{eq:theta_expansion} & Shift final-investor demand. \\
\(Z_t\) & Demand states, including flows, redemptions, and mandates & \eqref{eq:raw_demand}, \eqref{eq:raw_residual}, \eqref{eq:theta_expansion}, \eqref{eq:z_process} & Source of demand shocks. \\
\(\bar D(P_t,X_t,Z_t)\) & Raw final-investor demand function & \eqref{eq:raw_demand}, \eqref{eq:raw_residual}, \eqref{eq:normalized_supply_demand} & Determines how much final investors are willing to hold. \\
\(\bar q_t^E\) & Raw final-investor holdings & \eqref{eq:raw_demand}, \eqref{eq:raw_clearing} & First component of the raw market-clearing equation. \\
\(\bar S_t\) & Raw net supply & \eqref{eq:raw_clearing}, \eqref{eq:raw_residual}, \eqref{eq:normalized_supply_demand} & Total supply to be allocated across investors. \\
\(\bar q_t^A\) & Raw residual shares of risk absorbers & \eqref{eq:raw_clearing}, \eqref{eq:raw_residual}, \eqref{eq:theta} & Market-clearing residual in share units. \\
\(K_t^A\) & Dollar capital of the absorbing sector used for scaling & \eqref{eq:normalized_supply_demand}, \eqref{eq:theta} & Converts share exposure into exposure per dollar of absorbing capital. \\
\(s_t\) & Normalized supply & \eqref{eq:normalized_supply_demand}, \eqref{eq:theta}, \eqref{eq:theta_expansion} & Supply per unit of absorbing capital. \\
\(n_t(P_t,X_t,Z_t)\) & Normalized final demand & \eqref{eq:normalized_supply_demand}, \eqref{eq:theta}, \eqref{eq:theta_expansion} & Final demand per unit of absorbing capital. \\
\(\theta_t^{req}\) & Required residual exposure & \eqref{eq:theta}, \eqref{eq:theta_equilibrium}, \eqref{eq:theta_expansion} & Normalized exposure implied by market clearing. \\
\(\theta_t^A\) & Absorber controlled inventory state & \eqref{eq:theta_equilibrium}, \eqref{eq:fv_trading}, \eqref{eq:euler_general}, \eqref{eq:euler_quadratic} & State variable in the local HJB. \\
\(\theta_t\) & Common equilibrium exposure after imposing \(\theta_t^A=\theta_t^{req}\) & \eqref{eq:stock_flow}, \eqref{eq:inventory_decay} & Shorthand used after market clearing and the HJB state are matched. \\
\(u_t\) & Trading speed or inventory-adjustment speed & \eqref{eq:fv_trading}, \eqref{eq:trading_cost}, \eqref{eq:cost_function}, \eqref{eq:euler_general}, \eqref{eq:euler_quadratic}, \eqref{eq:stock_flow} & Control with \(\dd\theta_t^A=u_t\dd t\). \\
\(n_{P,t}\) & Local derivative of normalized dollar demand with respect to price & \eqref{eq:theta_expansion} & Enters the local price-response condition; its sign is imposed in normalized exposure units, not inferred from raw share-demand slope. \\
\(n_{X,t}\) & Local derivative of final demand with respect to characteristics & \eqref{eq:theta_expansion} & Describes how characteristic changes shift residual exposure. \\
\(n_{Z,t}\) & Local derivative of final demand with respect to demand states & \eqref{eq:theta_expansion} & Maps flows and redemptions into \(\theta_t\). \\
\(\dd P_t\), \(\dd X_t\), \(\dd Z_t\), \(\dd s_t\) & Changes in prices, characteristics, demand states, and supply & \eqref{eq:theta_expansion} & Decompose the sources of changes in \(\dd\theta_t^{req}\). \\
\addlinespace[3pt]
\multicolumn{4}{l}{\textit{Panel C. Exogenous State Variables}}\\
\addlinespace[2pt]
\(\mathcal Y_t\) or \(Y_t\) & State vector \((Z_t,K_t,\Sigma_t,\Phi_t,\Psi_t)\) & \eqref{eq:z_process}--\eqref{eq:sigma_process}; \(p_t=V_\theta(\theta_t^A,Y_t)\) & Summarizes demand, capital, risk, trading, and funding states. \\
\(\mathcal L^Y\) & Generator of \(Y_t\) & \eqref{eq:generator_exchange}, \eqref{eq:controlled_generator} & Acts on exogenous state variables and, in the baseline, has coefficients independent of \(\theta^A\). \\
\(\mathcal A^u\) & Controlled generator of \((\theta^A,Y)\) & \eqref{eq:controlled_generator}, \eqref{eq:euler_general}, \eqref{eq:euler_quadratic} & Gives the predictable drift of the inventory shadow value under \(\dd\theta_t^A=u_t\dd t\). \\
\(K_t\) & Normalized absorbing-capital or balance-sheet-capacity state & \eqref{eq:k_process}, \eqref{eq:capital_cost}, \eqref{eq:euler_quadratic} & Determines the tightness of capital constraints. \\
\(\Phi_t\) & Trading-cost or trading-capacity matrix & \eqref{eq:trading_cost}, \eqref{eq:cost_function}, \eqref{eq:euler_quadratic}, \eqref{eq:wedge_decomposition}, \eqref{eq:stock_flow} & Higher values represent more expensive trading. \\
\(\Psi_t\) & Financing or balance-sheet haircut matrix & \eqref{eq:capital_cost}; \(\mathcal Y_t\) & Governs the smooth component of financing costs. \\
\(Z_t\) shock terms & \(b_Z\), \(\sigma_Z\), \(W_t^Z\), \(J_Z\), and \(N_t\) & \eqref{eq:z_process} & Describe ordinary demand shocks and block imbalances. \\
\(b_K,\sigma_K,W_t^K\) & Drift and diffusion terms for \(K_t\) & \eqref{eq:k_process} & Describe changes in absorption capacity. \\
\(b_\Sigma,\sigma_\Sigma,W_t^\Sigma\) & Drift and diffusion terms for \(\Sigma_t\) & \eqref{eq:sigma_process} & Describe changes in risk conditions. \\
\addlinespace[3pt]
\multicolumn{4}{l}{\textit{Panel D. Cost Function and Optimization Variables}}\\
\addlinespace[2pt]
\(\gamma_t\) & Effective risk aversion or scarcity of risk-bearing capacity & \eqref{eq:inventory_cost}, \eqref{eq:cost_function}, \eqref{eq:euler_quadratic}, \eqref{eq:inventory_covariance}, \eqref{eq:inventory_risk_price} & Scales the price of inventory risk. \\
\(c_t(\theta,u,K)\) & Total instantaneous cost rate & \eqref{eq:absorber_objective}, \eqref{eq:cost_function}, \eqref{eq:euler_general} & Combines inventory, trading, and financing costs. \\
\(\ell_t(\theta,K)\) & Financing and capital cost & \eqref{eq:capital_cost}, \eqref{eq:static_absorption}, \eqref{eq:cost_function}, \eqref{eq:euler_quadratic}, \eqref{eq:wedge_decomposition} & Generates the balance-sheet wedge. \\
\(\eta_t\) & Strength of the nonlinear capital constraint & \eqref{eq:capital_cost} & Raises costs when required capital exceeds \(K_t\). \\
\(m_t\) & Capital-use or haircut weights & \eqref{eq:capital_cost} & Determines how much capacity each exposure consumes. \\
\(|\theta|\) & Absolute position vector & \eqref{eq:capital_cost} & Measures gross exposure for capital usage. \\
Excess-capital term & \(\max\{0,m_t^\top|\theta|-K_t\}\), the excess required capital above capacity & \eqref{eq:capital_cost} & Activates the nonlinear capital penalty. \\
\(\rho\) & Discount rate & \eqref{eq:absorber_objective}, \eqref{eq:euler_general}, \eqref{eq:euler_quadratic}, \eqref{eq:wedge_decomposition} & Trades off current compensation and future adjustment costs. \\
\(V(\theta,\mathcal Y)\) & Risk absorber's value function & Text definition; derives \(p_t\) & Provides the basis for the inventory shadow value. \\
\(p_t=V_\theta(\theta_t^A,Y_t)\) & Shadow value of inventory & \eqref{eq:euler_general}; \(p_t=\Phi_tu_t\) under quadratic adjustment costs & State variable for dynamic adjustment costs. \\
\(\nabla_\theta c_t\) & Marginal cost with respect to inventory & \eqref{eq:euler_general} & Current compensation required for holding inventory. \\
\(\nabla_u c_t\) & Marginal cost with respect to trading speed & \(p_t=\nabla_u c_t\) & Links trading speed to the shadow value. \\
\(\nabla_\theta\ell_t\) & Marginal financing or capital cost & \eqref{eq:static_absorption}, \eqref{eq:euler_quadratic}, \eqref{eq:wedge_decomposition} & Increases when capital constraints tighten. \\
\(\mathcal A^{u_t}p(\theta_t^A,Y_t)\) & Controlled-generator drift of the shadow value & \eqref{eq:euler_general}, \eqref{eq:euler_quadratic}, \eqref{eq:wedge_decomposition} & Converts expected changes in the inventory shadow value into current return compensation. \\
\(h_t\) & Additional hedging term & \eqref{eq:euler_quadratic} & Captures future demand, capital, or opportunity risks not explicitly summarized by the current state. \\
\(h_t^{wedge}\) & Hedging component inside the wedge & \eqref{eq:wedge_decomposition} & Part of the absorption wedge. \\
\addlinespace[3pt]
\multicolumn{4}{l}{\textit{Panel E. Empirical Projection Variables}}\\
\addlinespace[2pt]
\(\theta_t^{res}\) & Residual flow pressure & \eqref{eq:stock_flow}, \eqref{eq:inventory_decay} & New residual supply flow entering the absorber's inventory. \\
\(\E_t[\dd\theta_t^{res}/\dd t]\) & Expected future residual flow pressure & \eqref{eq:stock_flow} & Prices predictable future selling or residual supply. \\
\(\kappa_t\) & Absorption-capacity scarcity & \eqref{eq:stock_flow} & State variable measuring scarcity of market absorption capacity. \\
\(\theta_t\otimes\kappa_t\) & Interaction of inventory and scarcity & \eqref{eq:stock_flow} & Captures that the same inventory pressure is more expensive in scarce-capacity states. \\
\(A_t\) & Local coefficient on inventory & \eqref{eq:stock_flow} & Reduced-form price of residual inventory. \\
\(B_t\) & Local coefficient on expected residual flow & \eqref{eq:stock_flow} & Reduced-form price of predictable flow pressure. \\
\(C_t\) & Local coefficient on trading speed & \eqref{eq:stock_flow} & Reduced-form price of trading adjustment. \\
\(G_t\) & Local coefficient on scarcity & \eqref{eq:stock_flow} & Price of the scarcity state itself. \\
\(H_t\) & Local coefficient on inventory--scarcity interaction & \eqref{eq:stock_flow} & Core state-dependence prediction. \\
\(\varepsilon_t\) & Projection residual & \eqref{eq:stock_flow} & Component not captured by the local linear projection. \\
\(\Delta\) & Discrete time interval & \eqref{eq:inventory_decay} & Converts continuous-time dynamics into monthly or discrete data. \\
\(\delta_i\) & Inventory decay speed for asset \(i\) & \eqref{eq:inventory_decay} & Determines how quickly forced sales are absorbed. \\
\(o(\Delta)\) & Higher-order approximation term & \eqref{eq:inventory_decay} & Discrete-time approximation error. \\
\end{longtable}
\endgroup

\section{Additional Robustness Tables}

\begingroup
\captionsetup[table]{font=small,skip=3pt}
\setlength{\tabcolsep}{4.5pt}
\renewcommand{\arraystretch}{0.90}

\begin{table}[!htbp]
\centering
\begin{threeparttable}
\caption{Inventory-Decay Robustness}
\label{tab:rho_robustness}
\footnotesize
\begin{tabular}{lrrrr}
\toprule
Decay parameter & 1m & 3m & 6m & 12m \\
\midrule
\(\rho=0.50\) & 46.88 & 112.15 & 171.18 & 227.02 \\
 & (2.24) & (2.29) & (2.00) & (1.31) \\
\(\rho=0.75\) & 44.77 & 126.47 & 210.93 & 281.15 \\
 & (2.39) & (2.52) & (2.35) & (1.50) \\
\(\rho=0.85\) & 45.81 & 135.39 & 237.07 & 308.55 \\
 & (2.52) & (2.65) & (2.63) & (1.60) \\
\(\rho=0.90\) & 48.21 & 143.59 & 261.45 & 338.97 \\
 & (2.65) & (2.81) & (2.88) & (1.76) \\
\bottomrule
\end{tabular}
\begin{tablenotes}[flushleft]
\scriptsize
\item No-microcap Fama-MacBeth slopes. Coefficients are basis points; t-statistics are in parentheses.
\end{tablenotes}
\end{threeparttable}
\end{table}

\begin{table}[!htbp]
\centering
\begin{threeparttable}
\caption{Scarcity Decomposition}
\label{tab:scarcity_decomposition}
\footnotesize
\begin{tabular}{lrrrr}
\toprule
Variable & 1m & 3m & 6m & 12m \\
\midrule
\(\AbsInv\) rank & 17.03 & 48.08 & 75.02 & 91.45 \\
 & (2.44) & (2.93) & (2.45) & (1.59) \\
\(\AbsInv\times\) market illiquidity & -12.40 & -31.73 & -49.28 & -61.59 \\
 & (-2.43) & (-2.72) & (-2.31) & (-1.54) \\
\(\AbsInv\times\) VIX & 4.40 & 0.28 & -6.70 & -0.12 \\
 & (0.90) & (0.03) & (-0.41) & (-0.01) \\
\(\AbsInv\times\) NFCI & -6.01 & -18.18 & -34.18 & -44.82 \\
 & (-1.60) & (-2.03) & (-2.09) & (-1.53) \\
\(\AbsInv\times\) MF redemption stress & -1.72 & -0.88 & -6.30 & -30.69 \\
 & (-0.50) & (-0.12) & (-0.50) & (-1.24) \\
\(\AbsInv\) rank, excluding redemption from scarcity & 37.02 & 115.43 & 201.09 & 268.00 \\
 & (2.51) & (2.68) & (2.66) & (1.63) \\
\(\AbsInv\times\) nonflow scarcity & -4.61 & -25.24 & -57.73 & -57.96 \\
 & (-0.75) & (-1.62) & (-2.27) & (-1.51) \\
\bottomrule
\end{tabular}
\begin{tablenotes}[flushleft]
\scriptsize
\item No-microcap Fama-MacBeth slopes from specifications that decompose the scarcity proxy. VIX is the Cboe Volatility Index, NFCI is the National Financial Conditions Index, and MF denotes mutual fund. Coefficients are in basis points and t-statistics are in parentheses. The mixed signs are why the main text uses the composite scarcity split as the cleaner absorption-capacity diagnostic.
\end{tablenotes}
\end{threeparttable}
\end{table}

\begin{table}[!htbp]
\centering
\begin{threeparttable}
\caption{Nonparametric Scarcity-State Splits}
\label{tab:nonparametric_scarcity}
\footnotesize
\begin{tabular}{llrrr}
\toprule
State variable & State & 1m & 3m & 6m \\
\midrule
Composite scarcity & Low & 39.24 & 120.07 & 210.90 \\
 &  & (2.71) & (3.16) & (2.94) \\
Composite scarcity & High & 83.37 & 143.42 & 169.82 \\
 &  & (2.60) & (2.01) & (1.45) \\
Market illiquidity & Low & 48.16 & 111.60 & 179.88 \\
 &  & (1.17) & (2.14) & (1.98) \\
Market illiquidity & High & 88.54 & 190.75 & 253.62 \\
 &  & (4.07) & (4.07) & (4.30) \\
NFCI & Low & 50.32 & 95.21 & 137.24 \\
 &  & (1.22) & (2.64) & (2.11) \\
NFCI & High & 91.27 & 170.33 & 254.80 \\
 &  & (3.98) & (3.31) & (3.90) \\
MF redemption stress & Low & 23.90 & 105.39 & 137.50 \\
 &  & (0.64) & (3.03) & (2.45) \\
MF redemption stress & High & 92.22 & 114.47 & 199.31 \\
 &  & (2.42) & (1.72) & (1.52) \\
Nonflow scarcity & Low & 2.84 & 94.85 & 176.76 \\
 &  & (0.16) & (2.78) & (2.18) \\
Nonflow scarcity & High & 75.34 & 108.63 & 281.95 \\
 &  & (2.17) & (0.96) & (2.89) \\
\bottomrule
\end{tabular}
\begin{tablenotes}[flushleft]
\scriptsize
\item No-microcap Fama-MacBeth slopes of \(\AbsInv\) estimated separately in bottom- and top-tercile months for each state variable. Coefficients are in basis points and t-statistics are in parentheses. NFCI is the National Financial Conditions Index. The table is a nonlinear state diagnostic; it is not a claim that every component produces a monotone continuous interaction.
\end{tablenotes}
\end{threeparttable}
\end{table}

\begin{table}[!htbp]
\centering
\begin{threeparttable}
\caption{13F Absorber Decomposition}
\label{tab:absorber_decomposition}
\footnotesize
\begin{tabular}{lrr}
\toprule
Next-quarter 13F change & Coefficient & T-statistic \\
\midrule
All 13F institutions & -165.37 & -1.18 \\
High-turnover managers & -63.77 & -0.81 \\
Middle-turnover managers & -91.01 & -1.32 \\
Low-turnover managers & -10.59 & -0.87 \\
High minus low turnover & -53.18 & -0.64 \\
Bank typecode & -2.92 & -0.29 \\
Insurance typecode & -3.79 & -0.67 \\
Investment-company typecode & -3.53 & -1.80 \\
Independent-advisor typecode & -24.61 & -1.34 \\
Other or missing typecode & -130.52 & -1.12 \\
\bottomrule
\end{tabular}
\begin{tablenotes}[flushleft]
\scriptsize
\item No-microcap quarterly Fama-MacBeth slopes of next-quarter 13F dollar changes scaled by market equity on forced-sale pressure. Manager-turnover and typecode splits are diagnostic. Coefficients are in basis points.
\end{tablenotes}
\end{threeparttable}
\end{table}

\begin{table}[!htbp]
\centering
\begin{threeparttable}
\caption{Flow-Literature and Absorption Horse-Race Diagnostics}
\label{tab:horse_race}
\footnotesize
\begin{tabular}{lrrrr}
\toprule
Variable & 1m & 3m & 6m & 12m \\
\midrule
DistressSale, with standard flow controls & 29.82 & 68.42 & 96.84 & 172.23 \\
 & (1.53) & (1.84) & (1.70) & (1.67) \\
ExpectedSaleNext, out-of-sample & 106.86 & 244.10 & 326.65 & 408.18 \\
 & (2.54) & (3.07) & (3.16) & (2.49) \\
\(\AbsInv\times\) ThinBase, saturated horse race & 43.76 & 128.99 & 172.98 & 282.44 \\
 & (1.40) & (1.65) & (1.46) & (1.47) \\
\(\AbsInv\times\) Illiquidity, saturated horse race & 27.09 & 50.75 & 7.69 & -274.86 \\
 & (0.71) & (0.52) & (0.04) & (-1.01) \\
\(\AbsInv\times\) Scarcity, saturated horse race & -45.15 & -113.90 & -105.23 & -49.12 \\
 & (-1.22) & (-1.70) & (-0.81) & (-0.29) \\
\bottomrule
\end{tabular}
\begin{tablenotes}[flushleft]
\scriptsize
\item No-microcap diagnostic slopes. Saturated horse-race specifications include forced-sale, distress-sale, expected-sale, \(\AbsInv\), scarcity, thin-base, illiquidity interactions, and standard controls. Coefficients are in basis points.
\end{tablenotes}
\end{threeparttable}
\end{table}

\begin{table}[!htbp]
\centering
\begin{threeparttable}
\caption{Purchase-Side and Future-Fundamentals Diagnostics}
\label{tab:purchase_fundamentals}
\footnotesize
\begin{tabular}{lrrr}
\toprule
Diagnostic & 1m & 3m & 6m \\
\midrule
\multicolumn{4}{l}{Panel A: Contemporaneous price pressure} \\
ForcedSale rank, all months & -61.64 &  &  \\
 & (-6.20) &  &  \\
ForcedSale rank, high scarcity & -77.41 &  &  \\
 & (-4.05) &  &  \\
ForcedBuy rank, all months & -54.50 &  &  \\
 & (-4.26) &  &  \\
ForcedBuy rank, high scarcity & -79.14 &  &  \\
 & (-3.31) &  &  \\
\addlinespace[2pt]
\multicolumn{4}{l}{Panel B: Future returns after selling and buying pressure} \\
ForcedSale rank & 37.74 & 100.59 & 125.20 \\
 & (1.69) & (2.49) & (2.17) \\
ForcedBuy rank & 39.29 & 104.95 & 128.48 \\
 & (2.38) & (3.28) & (2.46) \\
\addlinespace[2pt]
\multicolumn{4}{l}{Panel C: Future returns with future fundamentals included as controls} \\
ForcedSale rank & 38.81 & 116.19 & 123.76 \\
 & (1.37) & (2.57) & (1.76) \\
\(\AbsInv\) rank & 33.03 & 66.63 & 78.35 \\
 & (1.56) & (1.29) & (0.89) \\
\bottomrule
\end{tabular}
\begin{tablenotes}[flushleft]
\scriptsize
\item No-microcap diagnostics. Panel A reports contemporaneous return slopes; empty entries indicate that the panel is not a future-return horizon. Panel B compares flow-induced selling and buying pressure with the same standard controls. Panel C adds look-ahead controls for future twelve-month changes in ROA, gross profitability, operating profitability, and sales growth. These controls are not available to investors at portfolio formation and are used only to diagnose whether the return relation is mechanically a future-fundamental improvement story. Coefficients are in basis points and t-statistics are in parentheses.
\end{tablenotes}
\end{threeparttable}
\end{table}

\begin{table}[!htbp]
\centering
\begin{threeparttable}
\caption{Cumulative Absorption Pressure by Size Tercile}
\label{tab:size_terciles}
\footnotesize
\begin{tabular}{lrrrr}
\toprule
No-microcap size tercile & 1m & 3m & 6m & 12m \\
\midrule
Small & -12.16 & 101.20 & 302.11 & 364.64 \\
 & (-0.20) & (0.66) & (1.17) & (0.71) \\
Medium & -23.16 & 129.02 & 364.22 & 611.96 \\
 & (-0.41) & (1.29) & (2.70) & (2.54) \\
Large & -206.05 & -200.16 & -333.17 & -625.18 \\
 & (-1.10) & (-1.06) & (-0.98) & (-1.04) \\
\bottomrule
\end{tabular}
\begin{tablenotes}[flushleft]
\scriptsize
\item Fama-MacBeth slopes for \(\AbsInv\) within NYSE-size terciles of the no-microcap sample. Coefficients are in basis points.
\end{tablenotes}
\end{threeparttable}
\end{table}

\begin{table}[!htbp]
\centering
\begin{threeparttable}
\caption{Turnover and Transaction-Cost Diagnostics}
\label{tab:turnover_costs}
\footnotesize
\begin{tabular}{lrr}
\toprule
Measure & Monthly mean & T-statistic \\
\midrule
Equal-weighted H-L gross return & 47.88 & 3.11 \\
Value-weighted H-L gross return & 24.03 & 1.50 \\
H-L one-way turnover & 17.18\% & 24.43 \\
Equal-weighted net return, 10 bps one-way cost & 46.54 & 3.00 \\
Equal-weighted net return, 25 bps one-way cost & 43.96 & 2.84 \\
Equal-weighted net return, 50 bps one-way cost & 39.67 & 2.56 \\
Equal-weighted net return, 100 bps one-way cost & 31.08 & 2.01 \\
Value-weighted net return, 25 bps one-way cost & 19.71 & 1.23 \\
Equal-weighted break-even one-way cost & 278.75 & \\
\bottomrule
\end{tabular}
\begin{tablenotes}[flushleft]
\scriptsize
\item No-microcap monthly high-minus-low absorption portfolio diagnostics. Return entries are basis points per month. Cost adjustments are illustrative.
\end{tablenotes}
\end{threeparttable}
\end{table}

\begin{table}[!htbp]
\centering
\begin{threeparttable}
\caption{Double Sorts by Mechanism Variables}
\label{tab:double_sorts}
\footnotesize
\begin{tabular}{lrrrr}
\toprule
Conditioning variable & 1m & 3m & 6m & 12m \\
\midrule
High ThinBase & 70.49 & 205.31 & 350.88 & 530.44 \\
 & (2.61) & (2.82) & (2.53) & (1.89) \\
High illiquidity & 63.88 & 175.42 & 296.37 & 474.10 \\
 & (2.44) & (2.51) & (2.17) & (1.72) \\
High scarcity & 76.21 & 195.41 & 326.79 & 524.75 \\
 & (2.64) & (2.89) & (2.46) & (1.89) \\
\bottomrule
\end{tabular}
\begin{tablenotes}[flushleft]
\scriptsize
\item Equal-weighted no-microcap high-minus-low absorption returns within high mechanism states. Coefficients are in basis points.
\end{tablenotes}
\end{threeparttable}
\end{table}

\begin{table}[!htbp]
\centering
\begin{threeparttable}
\caption{Strict Fixed-Effect Panel Specifications}
\label{tab:strict_fe}
\footnotesize
\begin{tabular}{lrr}
\toprule
Specification & ForcedSale coefficient & T-statistic \\
\midrule
Stock and month fixed effects & 64.81 & 1.79 \\
Stock and industry-month fixed effects & 62.28 & 2.21 \\
\bottomrule
\end{tabular}
\begin{tablenotes}[flushleft]
\scriptsize
\item The table reports one-month no-microcap panel estimates. The dependent variable is next-month return. Coefficients are in basis points.
\end{tablenotes}
\end{threeparttable}
\end{table}

\begin{table}[!htbp]
\centering
\begin{threeparttable}
\caption{Spearman Correlations Among Key Rank Variables}
\label{tab:correlations_appendix}
\footnotesize
\begin{tabular}{lrrrrrr}
\toprule
 & ForcedSale & \(\AbsInv\) & ThinBase & IO & Size & Illiq \\
\midrule
ForcedSale & 1.00 & 0.67 & -0.27 & 0.38 & -0.05 & -0.07 \\
\(\AbsInv\) & 0.67 & 1.00 & -0.38 & 0.57 & -0.09 & -0.08 \\
ThinBase & -0.27 & -0.38 & 1.00 & -0.62 & -0.54 & 0.69 \\
IO & 0.38 & 0.57 & -0.62 & 1.00 & -0.03 & -0.13 \\
Size & -0.05 & -0.09 & -0.54 & -0.03 & 1.00 & -0.86 \\
Illiq & -0.07 & -0.08 & 0.69 & -0.13 & -0.86 & 1.00 \\
\bottomrule
\end{tabular}
\begin{tablenotes}[flushleft]
\scriptsize
\item Spearman correlations among key monthly rank variables. IO is institutional ownership and Illiq is Amihud illiquidity.
\end{tablenotes}
\end{threeparttable}
\end{table}

\FloatBarrier
\endgroup

\end{document}